\documentclass{aa}
\usepackage[varg]{txfonts}

\usepackage{graphicx}
\usepackage{natbib}
\bibpunct{(}{)}{;}{a}{}{,} 
\usepackage{color}
\usepackage{float}
\usepackage{hyperref}
\usepackage{longtable}
\usepackage{rotating}
\usepackage{lscape}
\usepackage{soul}
\usepackage{subcaption}
\usepackage{adjustbox}

\usepackage{xcolor}

\newcommand{\lya}{\mbox{Ly$\alpha$}}

\newcommand{\flcgs}{\mbox{erg s$^{-1}$ cm$^{-2}$}}
\newcommand{\sbl}{\mbox{erg s$^{-1}$ cm$^{-2}$ arcsec$^{-2}$}}
\newcommand{\kms}{\mbox{km s$^{-1}$}}

\newcommand{\hii}{\mbox{H\,{\scshape ii}}}
\newcommand{\hi}{\mbox{H\,{\scshape i}}}
\newcommand{\mgii}{\mbox{Mg\,{\scshape ii}}}
\newcommand{\mgi}{\mbox{Mg\,{\scshape i}}}
\newcommand{\feii}{\mbox{Fe\,{\scshape ii}*}}
\newcommand{\oii}{\mbox{[O\,{\scshape ii}}]}
\newcommand{\oiii}{\mbox{[O\,{\scshape iii}}]}

\newcommand{\neiii}{\mbox{[Ne\,{\scshape iii}}]}
\newcommand{\cii}{\mbox{C\,{\scshape ii}}]}
\newcommand{\nv}{\mbox{N\,{\scshape v}}}

\newcommand{\udft}{\textsf{udf-10}}
\newcommand{\mosaic}{\textsf{mosaic}}

\begin{document}

\title{The MUSE Extremely Deep Field: a first panoramic view of an \mgii\ emitting intragroup medium\thanks{Based on observations made with ESO telescope at the La Silla Paranal Observatory under the large program 1101.A-0127}}

\author{Floriane Leclercq\inst{\ref{obsge}\thanks{e-mail: floriane.leclercq@unige.ch}}
\and Anne Verhamme\inst{\ref{obsge}} 
\and Benoit Epinat\inst{\ref{lam}}
\and Charlotte Simmonds\inst{\ref{obsge}}
\and Jorryt Matthee\inst{\ref{eth}}
\and Nicolas F. Bouché\inst{\ref{cral}}
\and Thibault Garel\inst{\ref{obsge},\ref{cral}}
\and Tanya Urrutia\inst{\ref{postdam}}
\and Lutz Wisotzki\inst{\ref{postdam}}
\and Johannes Zabl\inst{\ref{cral},\ref{hali}}
\and Roland Bacon\inst{\ref{cral}}
\and Valentina Abril-Melgarejo\inst{\ref{lam}}
\and Leindert Boogaard\inst{\ref{mpia},\ref{leiden}}
\and Jarle Brinchmann\inst{\ref{porto},\ref{leiden}}
\and Sebastiano Cantalupo\inst{\ref{milan},\ref{eth}}
\and Thierry Contini\inst{\ref{irap}}
\and Josephine Kerutt\inst{\ref{obsge}}
\and Haruka Kusakabe\inst{\ref{obsge}}
\and Michael Maseda\inst{\ref{leiden}}
\and Léo Michel-Dansac\inst{\ref{cral}}
\and Sowgat Muzahid\inst{\ref{pune},\ref{postdam}}
\and Themiya Nanayakkara\inst{\ref{swin}}
\and Johan Richard\inst{\ref{cral}}
\and Joop Schaye\inst{\ref{leiden}}
}

\institute{Observatoire de Genève, Universite de Genève, Chemin Pegasi 51, 1290 Versoix, Switzerland \label{obsge}
\and Aix Marseille Univ, CNRS, CNES, LAM, Marseille, France \label{lam}
\and Department of Physics, ETH Z\"{u}rich, Wolfgang-Pauli-Strasse 27, 8093 Z\"{u}rich, Switzerland \label{eth}
\and Univ Lyon, Univ Lyon1, Ens de Lyon, CNRS, Centre de Recherche Astrophysique de Lyon UMR5574, 69230 Saint-Genis-Laval, France \label{cral}
\and Institute for Computational Astrophysics and Department of Astronomy \& Physics,  Saint Mary's University, 923 Robie Street, Halifax, Nova Scotia, B3H 3C3, Canada \label{hali}
\and Leibniz-Institut für Astrophysik Potsdam (AIP), An der Sternwarte 16, 14482 Potsdam, Germany \label{postdam}
\and Max Planck Institute for Astronomy, K\"{o}nigstuhl 17, 69117 Heidelberg, Germany \label{mpia}
\and Leiden Observatory, Leiden University, PO Box 9513, NL-2300 RA Leiden, the Netherlands \label{leiden}
\and Instituto de Astrofísica e Ciências do Espaço, Universidade do Porto, CAUP, Rua das Estrelas, PT4150-762 Porto, Portugal \label{porto}
\and Dipartimento di Fisica G. Occhialini, Università degli Studi di Milano Bicocca, Piazza della Scienza 3, 20126 Milano, Italy\label{milan}
\and Institut de Recherche en Astrophysique et Planétologie (IRAP), Université de Toulouse, CNRS, UPS, CNES, 31400 Toulouse, France \label{irap}
\and IUCAA, Post Bag 04, Ganeshkhind, Pune-411007, India \label{pune}
\and Centre for Astrophysics and Supercomputing, Swinburne University of Technology, Hawthorn, VIC 3122, Australia \label{swin}
}



\abstract
{Using the exquisite MUSE eXtremely Deep Field data, we report the discovery of an \mgii\ emission nebula with an area above a 2$\sigma$ significance level of 1000 proper kpc$^2$, providing the first panoramic view of the spatial distribution of magnesium in the intragroup medium of a low mass group of five star-forming galaxies at $z=1.31$. The galaxy group members are separated by less than 50 physical kpc in projection and $\approx$120 km s$^{-1}$ in velocity space. The most massive galaxy has a stellar mass of 10$^{9.35}$ M$_\odot$ and shows an \mgii\ P-Cygni line profile indicating the presence of an outflow, which is consistent with the spatially resolved spectral analysis showing $\approx+$120 km s$^{-1}$ shift of the \mgii\ emission lines with respect to the systemic redshift. The other galaxies are less massive and only show \mgii\ in emission. The detected \mgii\ nebula has a maximal projected extent of $\approx$70 kpc including a low surface brightness ($\approx$ 2 $\times$ 10$^{-19}$ $\sbl$) gaseous bridge between two subgroups of galaxies. The presence of absorption features in the spectrum of a background galaxy located at an impact parameter of 19 kpc from the closest galaxy of the group indicates the presence of gas enriched in magnesium even beyond the detected nebula seen in emission, suggesting that we are observing the tip of a larger intragroup medium.
The observed \mgii\ velocity gradient suggests an overall rotation of the structure along the major axis of the most massive galaxy.
Our MUSE data also reveal extended \feii\ emission in the vicinity of the most massive galaxy, aligned with its minor axis and pointing towards a neighboring galaxy. Extended \oii\ emission is found around the galaxy group members and at the location of the \mgii\ bridge. Our results suggest that both tidal stripping effects from galaxy interactions and outflows are enriching the intragroup medium of this system.}


\maketitle

\section{Introduction}
\label{sec:1}

In the $\Lambda$CDM framework, theories and simulations predict that galaxies grow in mass and size by accreting cold gas and by merging with other galaxies. 
The environment is known to play a crucial role in shaping galaxy growth. Indeed, theoretical work indicates that tidal interactions and merging events can dramatically impact the properties of the galaxies and their circumgalactic medium (CGM, e.g. \citealt{Hani18}). 
Studying galaxy systems is therefore crucial for understanding the effect of environment on galaxy formation and evolution.

A large number of galaxy groups found in spectroscopic redshift surveys are reported in the literature (e.g. \citealt{Cucciati10,Iovino16,AbrilMel21}). 
Galaxy groups are generally referred to as the lower mass end (dark matter halo mass $\lesssim$10$^{13}$ M$_{\odot}$) of galaxy clusters, although the definition of groups in terms of halo mass is vague.
A commonly adopted definition of a galaxy group is an association of three or more galaxies with a physical distance $\Delta$r $\lesssim$ 500 kpc and a velocity separation $\Delta$v $\lesssim$ 500 $\kms$ (e.g. \citealt{Knobel09,Diener13}). 

Gas exchanges between the group members and their CGM as well as galaxy interactions shape the intragroup medium (IGrM). Therefore it is a great laboratory to study the effects of the local environment and gas flows on galaxy formation and evolution.
The IGrM is a multi-phase medium. At very low redshifts, it is mainly detected and studied through the Xray emission from its hot phase (e.g. \citealt{Lovisari15,Eckert17} and \citealt{Oppenheimer21} and associated papers for a review). Those studies are however limited to massive galaxy groups (dark matter, hereafter DM, halo mass > 10$^{13}$ M$_\odot$) due to the low X-ray flux at lower masses \citep{Lovisari21}.
Observations of neutral hydrogen in nearby galaxy groups reveal that the cold phase of the IGrM also shows extended structures \citep{Michel-Dansac10, Borthakur10,Serra13}.
At higher redshifts, most constraints on the associated IGrM come from absorption line studies in lines of sight probed by bright background objects. 
Several possibilities are proposed and discussed in the literature to explain the origin of the absorbing gas in systems with three or more galaxies:  the gas detected in absorption could be associated with the IGrM of the system \citep{N18}, the CGM of the group members \citep{B11,Fo19}, tidal tail remnants \citep{Ka10,Du20} or orbiting gas clouds \citep{Bi17} after galaxy interactions.

The advent of integral field unit (IFU) instruments is revolutionizing the study of the circum- and intragroup media by allowing their direct mapping in emission.
The unprecedented sensitivity of VLT/MUSE \citep{Ba10} recently allowed the detection of ionised and enriched intragroup nebula at $z\lesssim 0.7$. 
\cite{Ep18} reported the detection of a 150 kpc large $\oii$ nebula at $z\approx0.7$ surrounding a dozen galaxies with stellar masses around 10$^{10}$ M$_{\odot}$.
\cite{Ch19} reported a 100~kpc H$\alpha$ blob around a galaxy group with a dynamical mass of $\approx$ 3$\times$10$^{12}$ M$_{\odot}$ at $z\approx0.3$.
While \cite{Ep18} do not exclude a possible contribution from active galactic nucleus (AGN) outflows, both analyses favour a scenario where the observed metal enriched gas has been extracted from galaxies by tidal stripping forces.
Moreover, \cite{Johnson18} discovered six and \cite{Helton21} three spatially extended nebulae emitting in $\oiii$, $\oii$ and $\rm H \beta$ in $z\approx0.5$ galaxy groups hosting a quasar. Both studies suggest that nebulae are a signature of galaxy interactions in quasar host groups.

At higher redshift, the IGrM of star-forming galaxies is mainly detected in \hi\ Lyman $\alpha$ $\lambda 1215.67$ \AA\ ($\lya$) and generally referred to as $\lya$ blobs (LABs, e.g. \citealt{Steidel10,Caminha16,Ven17,Herenz20}). The $\lya$ line traces the neutral phase of the IGrM at $z>0.3$; at $z=0$ $\lya$ observations are extremely limited. In order to connect the ionized intragroup media observed at lower redshifts to those of high redshift galaxy systems, a tracer of the neutral gas phase observable at low $z$ is needed. The \mgii\ $\lambda\lambda$2796, 2803 (hereafter \mgii) doublet is a very good candidate. Indeed, because of the lower ionization potential of \mgi\ (7.6 eV) compared to \hi\ (13.6 eV), hydrogen gas is neutral when \mgii\ photons are emitted, i.e. \mgii\ traces the \hi\ gas. Because \mgii\ is a resonant line, we expect to observe extended \mgii\ emission, similar to $\lya$ (e.g. \citealt{W16,L17}).

The detection of extended \mgii\ emission was reported for the first time ten years ago by \cite{R11} and later by \cite{M13}, both using slit spectroscopy of galaxies at $z\approx0.7$ and 0.9, respectively. Statistical evidence for such \mgii\ halos at slightly higher redshift ($z\approx1.5$) was demonstrated by \cite{E12} by stacking long slit spectra. Their analyses were however limited by the slit aperture, preventing us from getting a complete view of the galaxy surroundings. By reporting non detections in narrow band (NB) images of five star-forming galaxies at $z\approx0.7$, \cite{RickardsVaught19} reinforced the idea that the detection of the extended \mgii\ gas is challenging. Recently, \cite{Burchett21} re-observed the \cite{R11} $z=0.7$ galaxy using the Keck Cosmic Webb Imager (Keck/KCWI, \citealt{Martin10kcwi}) and confirmed the detection of significant extended \mgii\ emission spanning over $\approx$40~kpc. They were able to perform the first detailed spatially resolved analysis of an \mgii\ halo and found that isotropic outflow models best fit the data. This detection has been followed by two similar discoveries (\citealt{Zabl21}, Wisotzki et al. in prep.) in the MUSE GAs FLOw and Wind (MEGAFLOW) survey and MUSE-Deep fields \citep{B17}, respectively.
\cite{Zabl21} reported the first discovery of an \mgii\ emission halo probed by a quasar sightline around a $z=0.702$ galaxy. Thanks to the 3D MUSE view on the CGM of this galaxy, the authors were able to compare the observations with toy models and found good consistency with a bi-conical outflow model.
These very recent discoveries highlight the fact that IFU instruments truly are game changers for the study of the circum- and intragroup medium of galaxies.

In this paper, we report the discovery of the first intragroup medium mapped in \mgii, detected in the MUSE eXtremely Deep Field (MXDF, Bacon et al. in prep. and see also \citealt{Bacon21}). The emitting nebula encompasses a group of five low mass galaxies (M$_*$< 10$^{9.4}$ M$_{\odot}$) at $z\approx1.31$ with a halo mass of $\approx$10$^{11.7}$ M$_{\odot}$. It extends over $\approx$70 kpc (physical) and unveils gaseous connections between galaxies. By taking advantage of the three-dimensional (3D) information provided by the MUSE data cubes, we study the kinematics of the nebula. Diffuse $\oii$~$\lambda\lambda3726,3729$ (hereafter \oii) and $\feii$ ($\lambda$2365, $\lambda$2396, $\lambda$2612, $\lambda$2626, hereafter \feii) emission are also detected and allow a comparison between the different gas phases of the IGrM. This first detection of an \mgii\ emitting intragroup nebula -- until now, only locally probed along absorption lines of sight -- allows us to shed light on the highly debated existence of an IGrM in low mass galaxy groups and on its origin.

The paper is organized as follows: we describe the MXDF observations, data reduction and catalog classification in Sect.~\ref{sec:2}. The group and the properties of the galaxy members studied in this paper are presented in Sect.~\ref{sec:3}. Section~\ref{sec:4} describes the detection and analysis of the $\mgii$ intragroup nebula. In Sect.~\ref{sec:5}, we investigate the $\oii$ and $\feii$ properties of the group and compare with $\mgii$. Then, in Sect.~\ref{sec:6}, we discuss the existence and origin of the intragroup medium and the mechanisms that make it shine. Section~\ref{sec:6} also provides a comparison with the literature.
Finally, we present our summary and conclusions in Sect.~\ref{sec:7}.\\

Throughout the paper, all magnitudes are expressed in the AB system and distances are in physical units, not comoving. We assume a flat $\Lambda$CDM cosmology with $\Omega_{\rm m}$ = 0.315 and
$\rm H_0$ = 67.4 $\kms$ Mpc$^{-1}$ \citep{Planck20}; in this framework, a 1\arcsec\ angular separation corresponds to 8.6 kpc proper at the redshift of the group ($z\approx1.31$).

\section{MXDF observations}
\label{sec:2}

Here we provide a summary of the data acquisition, data reduction and catalog building processes. More details can be found in \cite{Bacon21} and in the upcoming survey paper Bacon et al. (in prep).

The MXDF data were taken as part of the MUSE Guaranteed Time Observations program between August 2018 and January 2019 under photometric conditions. The observations were performed with the VLT GALACSI/AOF ground-layer adaptive optics system \citep{K16,Madec18}.
A single 140-hour pointing was completed after rejection of bad quality exposures. The MUSE field of view was rotated between each observation in order to reduce the systematics. The resulting quasi-circular field is more than 100 hours deep in the inner 31\arcsec\ radius and the exposure time decreases to $\approx$10 hours at a radius of 41\arcsec. The MXDF data partially overlaps with the MUSE 10-hour and 30-hour fields from \cite{B17} denoted \mosaic\ and \udft, respectively, as well as with the 1-hour MUSE-Wide fields (\citealt{Ur19}, Urrutia et al. in prep).

The data reduction is similar to the one described in \cite{B17}; it is based on the MUSE public pipeline \citep{W20} and includes some improvements resulting in less systematics and a better sky subtraction. 
We estimated the point spread function (PSF) using the \textsf{muse-psfrec} software \citep{Fusco20}. The final PSF is modeled using a Moffat function \citep{M69} with parameters of 0\farcs6 (0\farcs4) full width at half maximum (FWHM) and $\beta$ = 2.1 (1.8)  at the blue (red) end of the datacube. The datacube contains 157034 spatial pixels (spaxel) of 0\farcs2$\times$0\farcs2. The number of spectra matches the number of spaxels with a wavelength range of 4700 \AA\ to 9350 \AA\ divided into 3721 pixels of 1.25 \AA. 
The datacube also contains the estimated variance for each pixel.

The MXDF source catalog (Bacon et al. in prep) consists of (i) sources blindly detected and extracted using the \textsf{ORIGIN} software \citep{M20} which is designed to detect emission lines in 3D data sets and (ii) sources extracted using the \textsf{ODHIN} software \citep{bache2017} based on an HST catalog \citep{R15}, which performs source deblending in MUSE using high resolution HST images. 
The \textsf{ODHIN} approach is similar to \textsf{TDOSE} \citep{Schmidt2019} developed for the MUSE wide survey \citep{Ur19,Schmidt21} with the differences that it is non-parametric, it uses multiple broadband HST images, and it implements a regularization process to avoid noise amplification for very close sources.
The sources are then inspected by three experts and a final catalog of 733 sources with redshift and associated confidence is created. It includes 406 new spectroscopic redshift measurements with respect to the previous MUSE deep field catalog \citep{I17}.

\section{A group of galaxies at $z\approx1.3$}
\label{sec:3}

The \mgii\ nebula was discovered during the search for extended \mgii\ emission around galaxies in the MUSE \textit{Hubble} Ultra Deep Field (UDF, \citealt{B17}). A visual inspection, first in the \udft\ and then in the newly observed MXDF data (see Appendix~\ref{ap:4}), revealed at first glance \mgii\ emission offset from a galaxy and extended in one direction. By increasing the size of the search area and looking for neighbors, we found this object to be surrounded by four close galaxies. In this section we focus on the environment (Sect.~\ref{sec:31}), global properties and integrated spectral features of the group members (Sect.~\ref{sec:32} and Sect.~\ref{sec:33}, respectively). The \mgii\ intragroup nebula detected in the 5-member galaxy group is presented and analysed in the next section (Sect.~\ref{sec:4}).

\subsection{The galaxy group and its environment}
\label{sec:31}

\begin{figure*}
\centering
   \resizebox{\hsize}{!}{\includegraphics{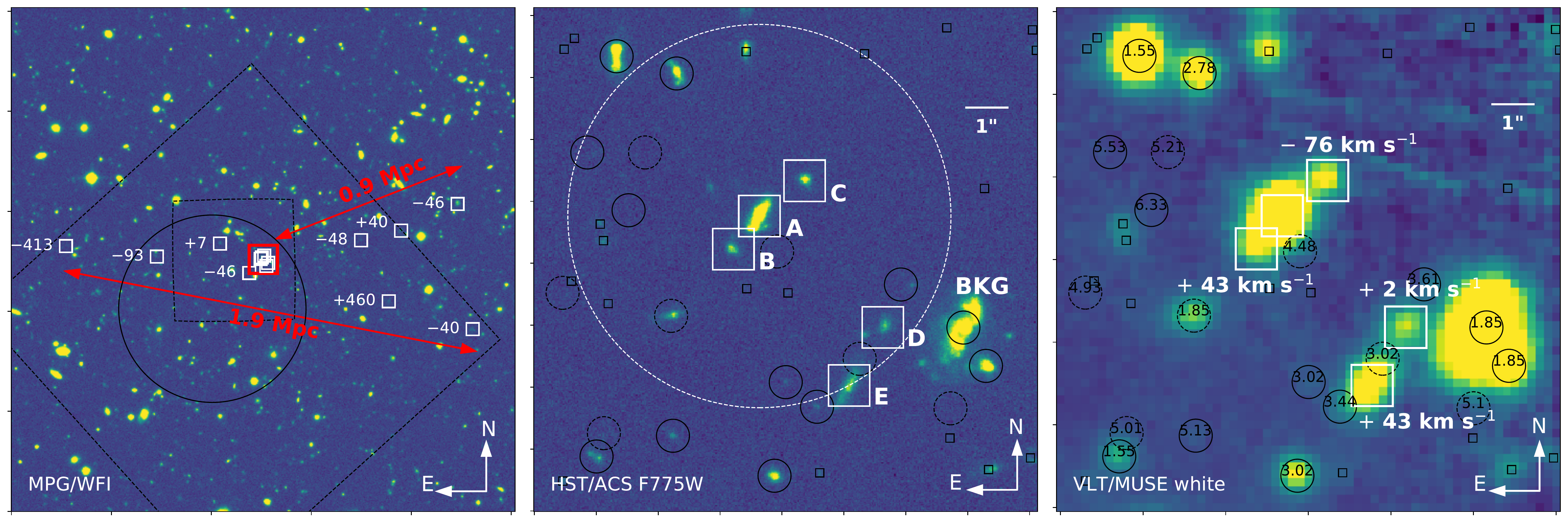}}
    \caption{\textit{Left:} The galaxies belonging to the same structure (see Sect.~\ref{sec:31}) as the galaxies embedded in the $z\approx1.31$ \mgii\ nebula (shown by the red square) are indicated by white squares. The velocity relative to the most massive galaxy of the group (galaxy A, see middle panel) is indicated for each galaxy in \kms. The structure spans $\approx$2 Mpc in projection and the galaxies appear well aligned. A $\approx$1 Mpc substructure extending beyond the MUSE Deep Field (big black dashed square) is also indicated. The MXDF and \udft\ fields are shown with the solid circle and small dashed square, respectively. The background image is an MPG(ESO2.2m)/WFI image \citep{Hildebrandt06}.
    \textit{Middle:} Zoom-in window of (12$\farcs$2 $\times$ 12$\farcs$2) on the five galaxies around which the \mgii\ nebula has been discovered. The galaxies are designated by letters where galaxy A is the most massive one (see Sect.~\ref{sec:32}), and the other letters (from B to E) are attributed to galaxies depending on their projected distance to galaxy A (from the closest to the furthest). The white circle is centered on galaxy A and has a radius of 40 kpc ($\approx$ 5\arcsec). The background image is the HST/ACS F775W image. The galaxies in the \citealt{R15}) HST catalog without a MUSE redshift are indicated by the small black squares. 
    The spectroscopic MUSE redshifts with confidence 1 and >1 (Bacon et al. in prep., see also \citealt{I17}) are indicated as black dashed and solid circles, respectively. The redshift values are indicated on the right panel. The label "BKG" points at a bright background galaxy acting as a sightline for absorption line studies of the \mgii\ nebula (see Sect.~\ref{sec:42}).
    \textit{Right:} Same spatial window as the middle panel showing the MUSE white light image of the group (white squares) and its surrounding. The positions of the galaxies B, C, D and E relative to galaxy A in velocity space are indicated in white. The redshift measurement procedure of the group members is described in Sect.~\ref{sec:33}. 
    }
    \label{fig:group_presentation}
\end{figure*}

The group we are studying here consists of five galaxies at redshift $z\approx1.31$.
Using a Friends-Of-Friends algorithm \citep{Huchra82} connecting galaxies separated by less than 450 kpc and 500 km s$^{-1}$ in the MXDF, UDF and MUSE-Wide catalogs (Bacon et al. in prep, \citealt{Ur19} and Urrutia et al. in prep), we found that this group is at the center of a larger structure of 14 galaxies with spectroscopically confirmed redshifts (with high confidence, ZCONF>1 in the MUSE catalogs except for one object, see below) and separated by less than 460 $\kms$ in redshift space.
This structure spans the MUSE UDF \mosaic\ field over 1.9 Mpc in projection (see Fig.~\ref{fig:group_presentation}, left panel). Three galaxies aligned in projection with the 5-galaxy group form a 0.9 Mpc long substructure extending in the MUSE-Wide fields, i.e. beyond the MUSE deep fields.

For the rest of the paper we will only focus on the subgroup of five galaxies located at the center of the larger structure (area within the red square in the first panel of Fig.~\ref{fig:group_presentation} and shown in the two other panels) because they are the galaxies embedded in the newly detected \mgii\ nebula. 
In order to ease the reading, the five galaxies are designated by letters\footnote{The MUSE identifiers in \cite{I17} are \#32, \#121, \#77 and \#65 for the galaxies A,C,D and E, respectively. The galaxy B is a newly detected source which is in the MXDF catalog only (Bacon et al. in prep) and has the identifier \#8493.} where galaxy A is the most massive one (see Sect.~\ref{sec:32}). The other letters (from B to E) are attributed to galaxies depending on their projected distance to galaxy A (from the closest to the furthest, see middle panel of Fig.~\ref{fig:group_presentation}). We note that galaxy B has a lower redshift confidence level (ZCONF=1) than the other group members in the MUSE UDF catalog (Bacon et al. in prep). Based on a careful analysis of the \oii\ NB image (see Sect.~\ref{sec:51}) -- where \oii\ emission is clearly detected at the position of galaxy B when setting low flux cuts -- and the rather well constrained photometric redshift (1.25 < $z_{\rm BPZ}$ < 1.47), we are confident about the redshift of this source and therefore about its group membership.

The group members are located within $\approx$ 50 kpc in projection and 120 $\kms$ in redshift space (middle and right panels of Fig.~\ref{fig:group_presentation}). The projected distances and velocity offsets from galaxy A are (9 kpc, $+$43 $\kms$), (13 kpc, $-$76 $\kms$), (37 kpc, $+$2 $\kms$) and (41 kpc, $+$43 $\kms$) for galaxies B, C, D and E, respectively (see right panel of Fig.~\ref{fig:group_presentation} and Table~\ref{tab:gal_prop}). Galaxies B and C are aligned with the major axis of galaxy A. Similarly, galaxy D is aligned with the major axis of galaxy E. 

A virial mass of $\approx10^{11}$ M$_\odot$ was estimated from the velocity dispersion of the five members using the gapper method \citep{Beers90,Cucciati10,AbrilMel21}. The dynamical halo mass becomes $\approx10^{12.3}$ M$_\odot$ when considering the 14 galaxies of the larger structure. The corresponding virial radius is $\approx$ 50 kpc (Eq.~1 of \citealt{Lemaux12}) for the 5-member group ($\approx 150$ kpc when considering the whole structure). 
Alternatively, using the stellar to halo mass relation from \cite{Behroozi19} at the redshift of the group, we calculated a DM halo mass of 10$^{11.5}$ M$_{\odot}$ for the central galaxy A (with 0.3 dex of uncertainties) and a virial radius of $\approx90$ kpc.
Those two methods therefore indicate that the five group members reside inside the virial radius of the group and galaxy A.

Galaxy A is also the brightest galaxy of the larger structure in the UDF ($\approx$ 24.5 mag in the F775W filter) and therefore corresponds to the dominant "brightest group galaxy" (BGG) near the halo center where the Xray intensity peak tracing the IGrM is usually observed in Xray studies \citep{Oppenheimer21}. The BGGs have been found to have disk-like morphologies \citep{Moffett16} like it seems to be the case for galaxy A given its elongated shape (Sect.~\ref{sec:32}). 
We note that one galaxy located at $\approx$1 Mpc outside the MUSE deep field and at the outskirt of the detected structure is brighter ($\approx$ 23 mag in the F775W filter) with a stellar mass of $\approx10^{10.6}$ M$_\odot$ \citep{Ur19}.

The middle panel of Fig.~\ref{fig:group_presentation} shows the F775W HST/ACS image of the group and its close environment in a $\approx$100 $\times$ 100 kpc$^2$ ($12\farcs2\times12\farcs2$) window. The known photometric redshift ranges (BPZ) from the \cite{R15} catalog are indicated for objects without MUSE spetroscopic redshifts. Galaxies with MUSE redshifts (Bacon et al. in prep., see also \citealt{I17}) are shown as solid or dotted black circles, depending on the redshift confidence level (middle and right panels of Fig.~\ref{fig:group_presentation}). The systemic redshifts of the five group members have been measured for this analysis in Sect.~\ref{sec:33}.

A bright background galaxy ($m_{\rm AB,F775W}=23.4$), referred to as "BKG", spectroscopically confirmed to be at $z\approx1.8463$ (MUSE ID \#18 in \citealt{I17}) is located at an impact parameter of 2$\farcs$2 (or 19 kpc) from galaxy D, i.e. within the virial radius of galaxy A.
This galaxy produces a strong continuum in the MUSE data and thus appears like an interesting sight line for the analysis of the \mgii\ nebula (see Sect.~\ref{sec:42}) in absorption. 
No other background projected neighbours, inside a 200 kpc radius around galaxy A, have strong enough detected continua in the MUSE data to allow the detection of absorption features.

\subsection{Global properties of the galaxies}
\label{sec:32}

The stellar masses and star formation rates (SFR) of the group members were estimated by fitting their spectral energy distribution (SED) from HST photometry (F225W, F275W, F336W, F435W, F606W, F775W, F850LP, F105W, F125W, F140W, F160W, \citealt{R15}) using \textsf{MAGPHYS} \citep{DaCunha08,DaCunha15}. This version of \textsf{MAGPHYS} uses a \cite{Ch03} initial mass function (IMF), \cite{BC03} stellar models and \cite{C00} dust attenuation. Moreover, the star formation histories (SFH) are considered exponential with random bursts superimposed. 
The resulting properties of the group members are given in Table~\ref{tab:gal_prop}. With a highest stellar mass of $\log M_{*}/ \rm  M_{\odot}=9.35_{-0.14}^{+0.16}$, the galaxies of this group are rather low mass objects. This is in contrast with previous studies where the systems embedded in ionized nebulae usually host several galaxies with stellar masses higher than 10$^{10} \rm \ M_{\odot}$ (e.g. \citealt{Ep18,Helton21}). Despite the large uncertainties on the SFR values of the less massive galaxies as well as the few constraints on the so-called "star formation main sequence" for low mass galaxies at $z = 1.3$, the five galaxies can be considered as normal star forming galaxies \citep{W14,Boogaard18}.

The two most massive galaxies (A and E) are seen edge-on with projected axis ratios of $q =$ 0.329$\pm$0.005 and $q =$ 0.247$\pm$0.015, respectively (corresponding to inclinations of 71 and 76 degrees, respectively, when considering a flat disk model) and a similar position angle PA $\simeq$ $-$27$^\circ$ (measurements from \citealt{VDW12} in the WFC3/F105W band, see Table~\ref{tab:gal_prop}). The three other galaxies have rounder shapes with $q$ $\gtrsim$ 0.7 and smaller sizes.

Regarding possible AGN contamination, we refer to \citealt{F18} (their Sect.~3.3). 
By cross-matching their \mgii\ sample (which includes four out of five of our group members) with the 7 Ms Source Catalogs of the \textit{Chandra} Deep Field South Survey \citep{Luo17}, the authors found an X-ray counterpart for only one source, which is not one of the galaxies analysed in this paper. We do not find any X-ray counterpart for galaxy B (not in \citealt{F18}). The faintness and low stellar masses of the galaxies (Table~\ref{tab:gal_prop}), the reasonable line widths, and the lack of typical AGN lines like \nv~$\lambda$3425, also indicate that the presence of an AGN in those galaxies is unlikely. Nonetheless, the presence of low luminosity or heavily obscured AGN can not be completely excluded.

\begin{table*}
    \centering
    \captionof{table}{Physical properties of the five group members}
    \def\arraystretch{1.5}
    \setlength{\tabcolsep}{5pt}
    \begin{tabular}{ccccccccccc}
    \hline
    ID & MID & $z_{\rm syst}$ & M$_{\rm UV}$ & log$_{10}$(M$_{*}$) & log$_{10}$(SFR) & $\Delta v_{\rm A}$& $\Delta p_{\rm A}$& $q$ & PA & EW$_{2796}$ \\
       &     &              & [mag]          & [M$_{\odot}$]      & [M$_{\odot}$ yr$^{-1}$] & [km s$^{-1}$] & [kpc] &   & [$^\circ$]  & [\AA] \\
    \hline
    \hline
    A & 32 & 1.30661 & $-$19.09$\pm$0.04 & 9.35$_{-0.14}^{+0.16}$ & 0.5$_{-0.06}^{+0.2}$ &  0   &  0&  0.329$\pm$0.005 & $-$26.5$\pm$0.4  & 1.5$\pm$0.2\\
    B & 8493 & 1.30695 & $-$16.71$\pm$0.17 & 8.49$_{-0.1}^{+0.08}$ & $-$0.48$_{-0.06}^{+0.08}$ & $+$43 & 9.2& 0.543$\pm$0.038 & 52$\pm$6 & 4.9$\pm$0.9\\
    C & 121 & 1.30603 & $-$17.09$\pm$0.11 & 8.55$_{-0.05}^{+0.1}$ & $-$0.55$_{-0.02}^{+0.07}$ & $-$76    & 13.0 & 0.559$\pm$0.032 & 54$\pm$6 & 1.3$\pm$0.2\\
    D & 77 & 1.30663 & $-$17.49$\pm$0.16 & 8.76$_{-0.06}^{+0.14}$ & 0.24$_{-0.32}^{+0.18}$ &  $+$2 &  37.3   & 0.644$\pm$0.038 & $-$12$\pm$6& 1.7$\pm$0.4\\
    E & 65 & 1.30694 & $-$17.85$\pm$0.11  & 8.94$_{-0.07}^{+0.14}$ & $-$0.07$_{-0.01}^{+0.1}$ &   $+$43 &   41.2   & 0.247$\pm$0.015 & $-$28$\pm$1  & 1.7$\pm$0.4\\
    \hline
    \end{tabular}
    \smallskip
    
    \small{\textbf{Notes.} ID: label of the galaxy group members as shown in Fig.~\ref{fig:group_presentation} (middle panel). MID: MUSE identifier (\citealt{I17}, Bacon et al. in prep.). z$_{\rm syst}$: systemic redshift as measured in Sect.~\ref{sec:33}. M$_{\rm UV}$: Absolute UV magnitude close to 1500 $\AA$ restframe measured in the F336W HST band \citep{R15}. log$_{10}$(M$_{*}$): logarithm of the stellar mass in M$_{\odot}$. SFR: star formation rate in M$_{\odot}$ yr$^{-1}$. The parameters M$_{*}$ and SFR are based on SED fitting (Sect.~\ref{sec:32}). $\Delta v_{\rm A}$: velocity offset from galaxy A in km s$^{-1}$. $\Delta p_{\rm A}$ projected distance from galaxy A in kpc. $q$: projected axis ratio. PA: position angle in degree. The galaxy morphological parameters q and PA are from \cite{VDW12} and measured in the WFC3/F105W band. EW$_{2796}$: rest frame equivalent width of the \mgii\ $\lambda$2796 line in \AA\ as measured by \textsf{pyplatefit} (Bacon et al. in prep., Sect~\ref{sec:33}.}
    \label{tab:gal_prop}
    
\end{table*}

\subsection{Integrated spectra}
\label{sec:33}

\begin{figure*}
\centering
   \resizebox{\hsize}{!}{\includegraphics{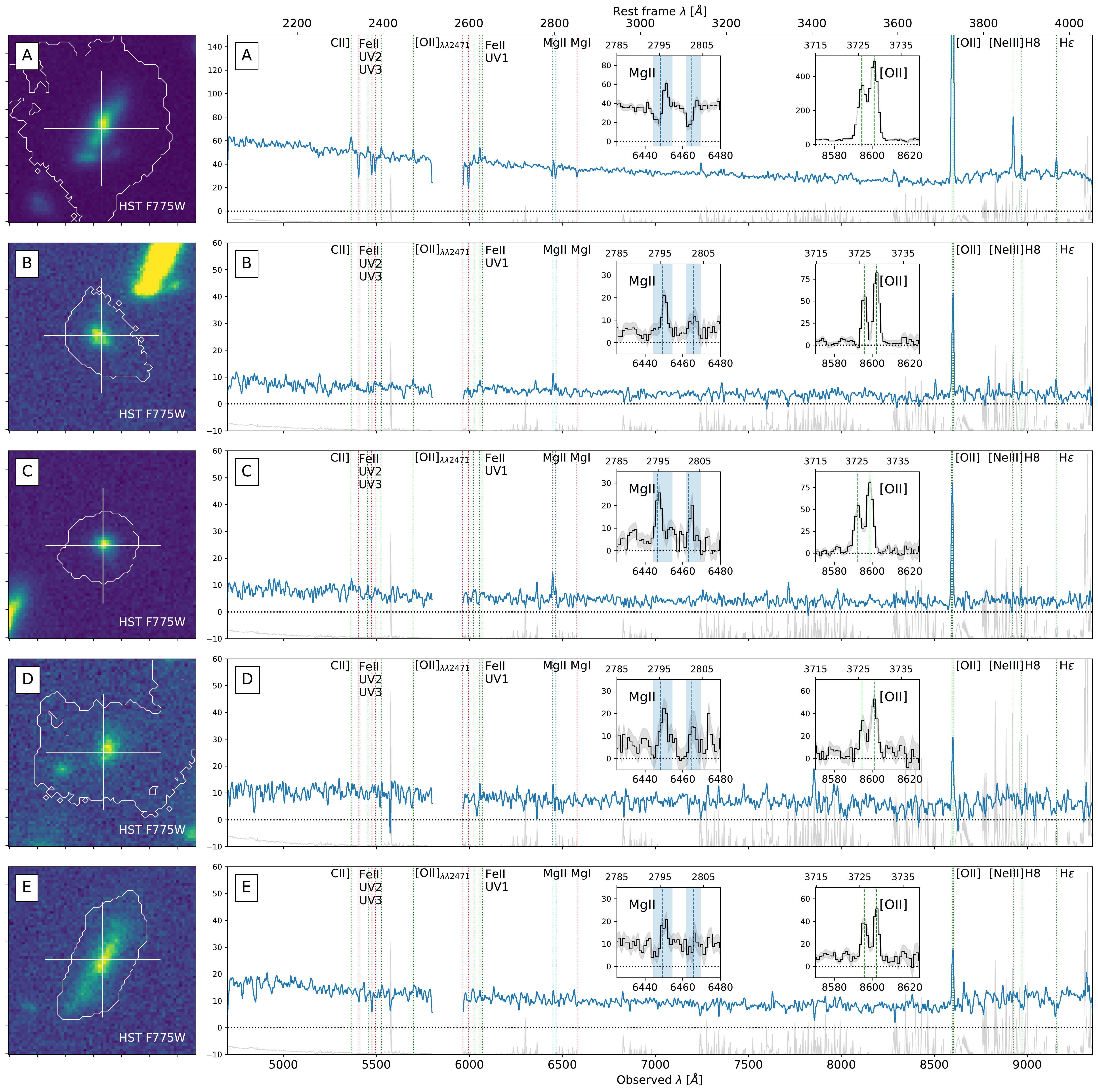}}
    \caption{\textit{Left:} $2\arcsec\times2\arcsec$ HST/ACS F775W images of the five group members. The central white cross and contour indicate the HST coordinates and segmentation map from \cite{R15}. \textit{Right:} Integrated spectrum (units of 10$^{-20}$ erg s$^{-1}$ cm$^{-2}$ \AA$^{-1}$) of the galaxy shown in the corresponding left panel extracted within the HST segmentation map contour using the {\textsf{ODHIN}} software (\citealt{bache2017}, see Sect. 3.3) and smoothed with a 7 \AA\  FWHM Gaussian function. The positions of several emission (absorption) lines are indicated by vertical dotted green (red) lines. The grey unsmoothed spectra show the 1$\sigma$ uncertainties which have been offset for readability. The unsmoothed \mgii\ and \oii\ lines (vertical dashed lines) are shown in the insets. The vertical blue shaded areas indicate the spectral widths used to construct the \mgii\ NB image shown in Fig.~\ref{fig:nb_mgii}a.}
    \label{fig:integ_sp}
\end{figure*}

Figure~\ref{fig:integ_sp} displays a zoom-in HST/ACS view in the F775W filter (first column) and the integrated spectrum (second column) of each group member. We display the {\textsf{ODHIN}} spectrum (see Sect.~\ref{sec:2}) which makes use of the higher spatial resolution of the HST images to perform deblending of sources in the MUSE cubes. The deblending feature of the {\textsf{ODHIN}} software is particularly needed here as the emission from the [A,B,C] and [D, E] galaxies are overlapping at MUSE resolution (see Fig.~\ref{fig:nb_mgii}a). The corresponding integrated error spectra are shown in grey in each panel and errors are directly indicated on the spectra (in grey) in the insets.

The systemic redshifts of the group members are calculated from those spectra by fitting the \oii\ doublet, which is the line with the highest signal to noise (S/N) for all sources. We fit the doublet using two Gaussian functions with fixed peak separation and considering the variance of the MUSE cube. The errors on the fit parameters are estimated using bootstrapping; 1000 realizations of the line doublet are generated using the error spectrum assuming that the errors are normally distributed around the observed flux values.
The resulting errors on the systemic redshift measurements (given in Table~\ref{tab:gal_prop}) are of the order of 10$^{-5}$.

All galaxies show a stellar continuum with clear \oii\ and \mgii\ emission (see inset panels in Fig.~\ref{fig:integ_sp}). 
We note that both \mgii\ and \oii\ doublets are impacted by significant skylines.
The spectrum of the most massive galaxy (A) displays other emission lines like \neiii, H8, H$\epsilon$ as well as \cii~$\lambda$2326 and \oii~$\lambda\lambda2470,2471$. 
Moreover, the Balmer absorption is detected on the red side of the \oii\ line. The \mgii\ line of galaxy A shows a P-Cygni profile and a detection of the \mgi~$\lambda$2852 absorption line. 
At the redshift of the group, the UV transitions \feii ($\lambda$2365, $\lambda$2396, $\lambda$2612, $\lambda$2626) are observable with MUSE; most of the corresponding emission and absorption lines are detected in the spectrum of galaxy A (top right panel of Fig.~\ref{fig:integ_sp}). 
Using the shallower \mosaic\ and \udft\ MUSE datacubes, \cite{F18} and \cite{Finley17} respectively studied the $\mgii$ and $\feii$ emission of star-forming galaxies. Both studies classified the $\mgii$ doublet of galaxy A as a P-Cygni profile. In \cite{F18}, galaxy C is classified as an $\mgii$ emitter and galaxy E as a non-detection. The redshift confidence level of galaxy D was not high enough to be included in their sample and, as mentioned before, galaxy B is a newly detected galaxy which thus was not in their sample either.

The \mgii\ rest frame equivalent widths (EW) of the \mgii~$\lambda$2796 line calculated using \textsf{pyplatefit} in Bacon et al. in prep. (see also \citealt{Bacon21}) are given in Table~\ref{tab:gal_prop}. The \textsf{Pyplatefit} software performs a stellar continuum fit using a simple population model \citep{Brinchmann2013} and fits the emission and absorption lines after subtracting the continuum. 
The intrinsic \mgii~$\lambda\lambda$2796,2803 absorption in the spectra of O/B stars is expected to be very weak ($\approx$0.6 \AA\ for population ages up to a few 100 Myrs, \citealt{H18}) and therefore can not explained the observed absorption. The \mgii\ absorption is most likely dominated by continuum photons pumping by Mg$^+$ ions along the line of sight, followed by isotropic re-emission out of line of sight, as a result of resonant scattering. This leads to a net absorption in front of the continuum sources, and possibly large scale diffuse emission that fills the absorption troughs. This effect is called emission infill (e.g., \citealt{Prochaska2011,ScarlataPanagia2015,Zhu2015,Finley17}).
The EW values reported in Table~\ref{tab:gal_prop} are not corrected for this effect. We also do not attempt to correct our \mgii\ observations for emission infill, neither the spectra nor the narrow bands (Sect.~\ref{sec:41}) because it would inevitably imply making the astrophysical assumption that absorbing and emitting regions are spatially different and can be separated. Since the physical meaning of any simple correction is limited, we present the full complexity of these \mgii\ data and highlight the novelty of the detection of such diffuse large-scale structures. Our non-corrected measurements enable (i) direct and simple comparisons with other studies in terms of detection limits, observed flux and spatial variations, and (ii) the reproducibility of our results and measurements. Finally, we note that the absorption correction has an effect at galaxy scale only, so our conclusions about the large scale intragroup medium are not impacted by the non-correction.

To sum up, we are looking at a compact group of five low mass members located at the center of a larger structure at high redshift. Taking advantage of the deep MXDF data, we analyse the \mgii\ properties of the IGrM in the next section (Sect.~\ref{sec:4}) and its \oii\ and \feii\ properties in Sect.~\ref{sec:5}.

\section{The \mgii\ intragroup nebula}
\label{sec:4}

\subsection{\mgii\ narrow-band image}
\label{sec:41}

The \mgii\ structure was extracted in a subcube of $12\farcs2\times12\farcs2$ ($\approx$100 $\times$ 100 kpc$^2$ at $z\simeq$ 1.31) centered on the coordinates RA = 53$^\circ$09\arcmin25\arcsec, Dec = $-$27$^\circ$46\arcmin44\arcsec. We estimated the continuum by performing a spectral median filtering on the subcube using a wide spectral window of 200 spectral pixels (250 $\AA$). After subtracting this continuum-only cube from the original one, we obtained an emission lines only cube from which we optimally created the \mgii\ NB image of the group as follows. 

In order to encompass all the \mgii\ emitting flux from the five galaxies of the group while limiting the noise, we adopted the following method: (i) the \mgii\ line is extracted by integrating inside a circular aperture including the five group members, centered on the continuum-subtracted subcube, which maximizes the integrated flux of the \mgii\ $\lambda$2796 line ($r=$ 3.2\arcsec), (ii) the NB image of the $\lambda$2796 \mgii\ line is created by summing the continuum-subtracted subcube over the wavelength range 6445.0$-$6453.75 $\AA$ ($\approx$ 400 \kms) delimiting the line; the borders of the line are determined by wavelengths for which the flux reaches zero. The same procedure is applied to extract the $\lambda$2803 line of the \mgii\ doublet (6462.5$-$6468.75 $\AA$, $\approx$ 290 \kms). The total \mgii\ emission NB image is finally obtained by adding the $\lambda$2796 and $\lambda$2803 NB images. A broader band \mgii\ image is presented in Fig.~\ref{fig:nb-vs-bb} and confirms that our method provides an S/N-optimized NB image without missing flux (see the residuals image in the right panel of Fig.~\ref{fig:nb-vs-bb}).

\begin{figure*}
\begin{minipage}{0.64\textwidth}\centering
   \resizebox{\hsize}{!}{\includegraphics{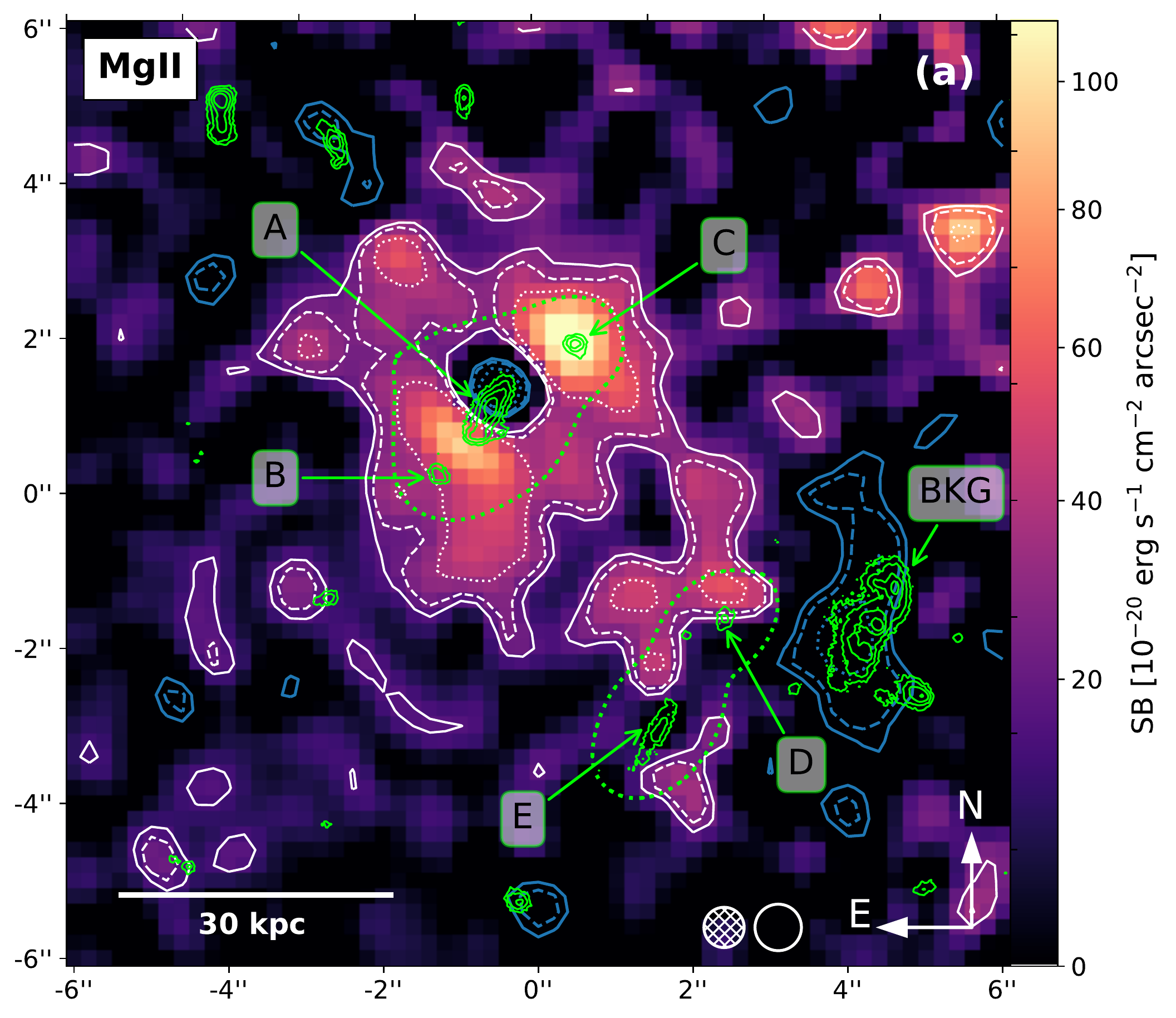}}
\end{minipage}
\hspace{0.1cm}
\begin{minipage}{0.36\textwidth}
    \begin{subfigure}[t]{0.95\linewidth}
    \vspace{0.1cm}
        \includegraphics[width=\textwidth]{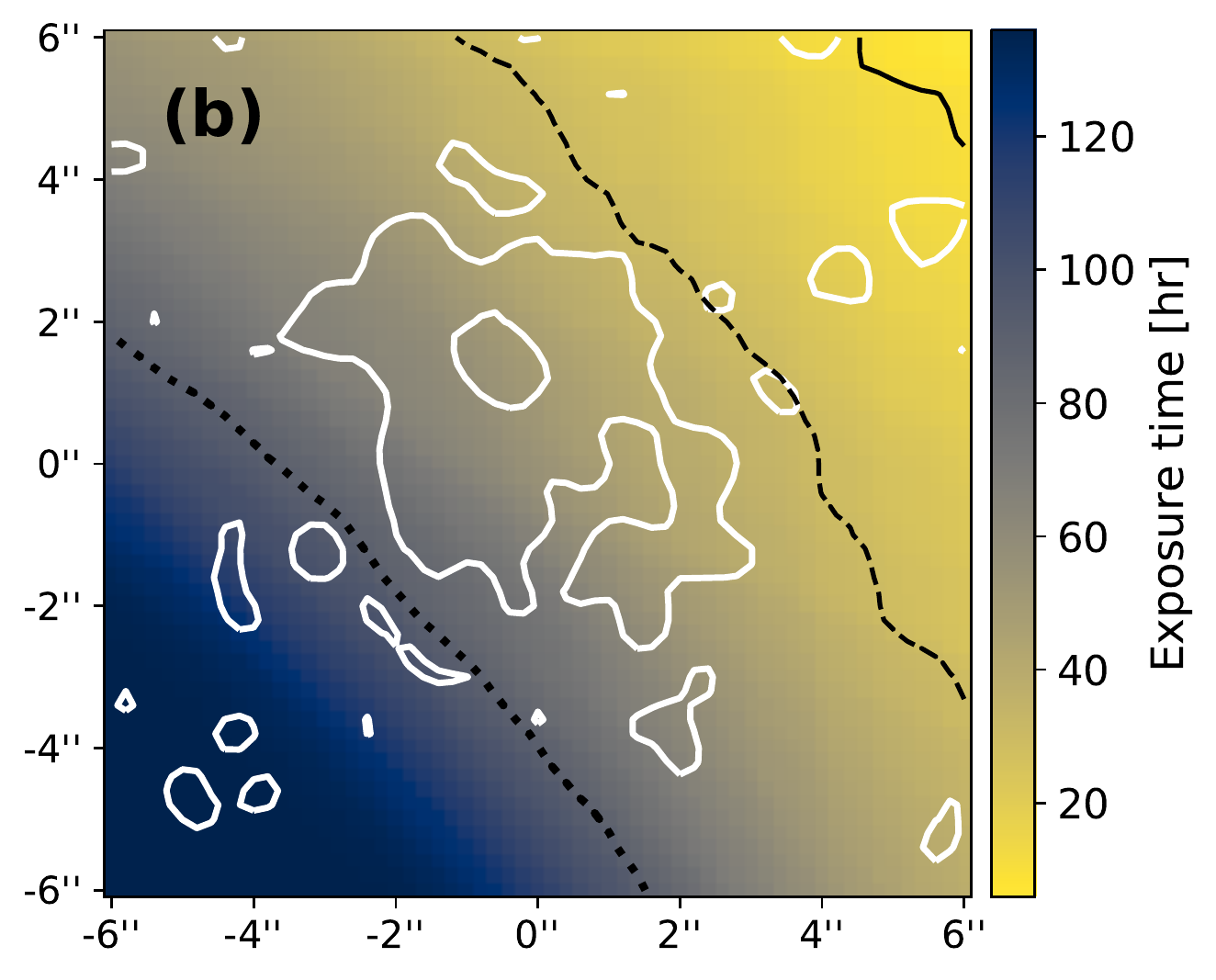}
    \end{subfigure} \\
    \begin{subfigure}[b]{0.95\linewidth}
        \vspace{0.1cm}
        \includegraphics[width=\textwidth]{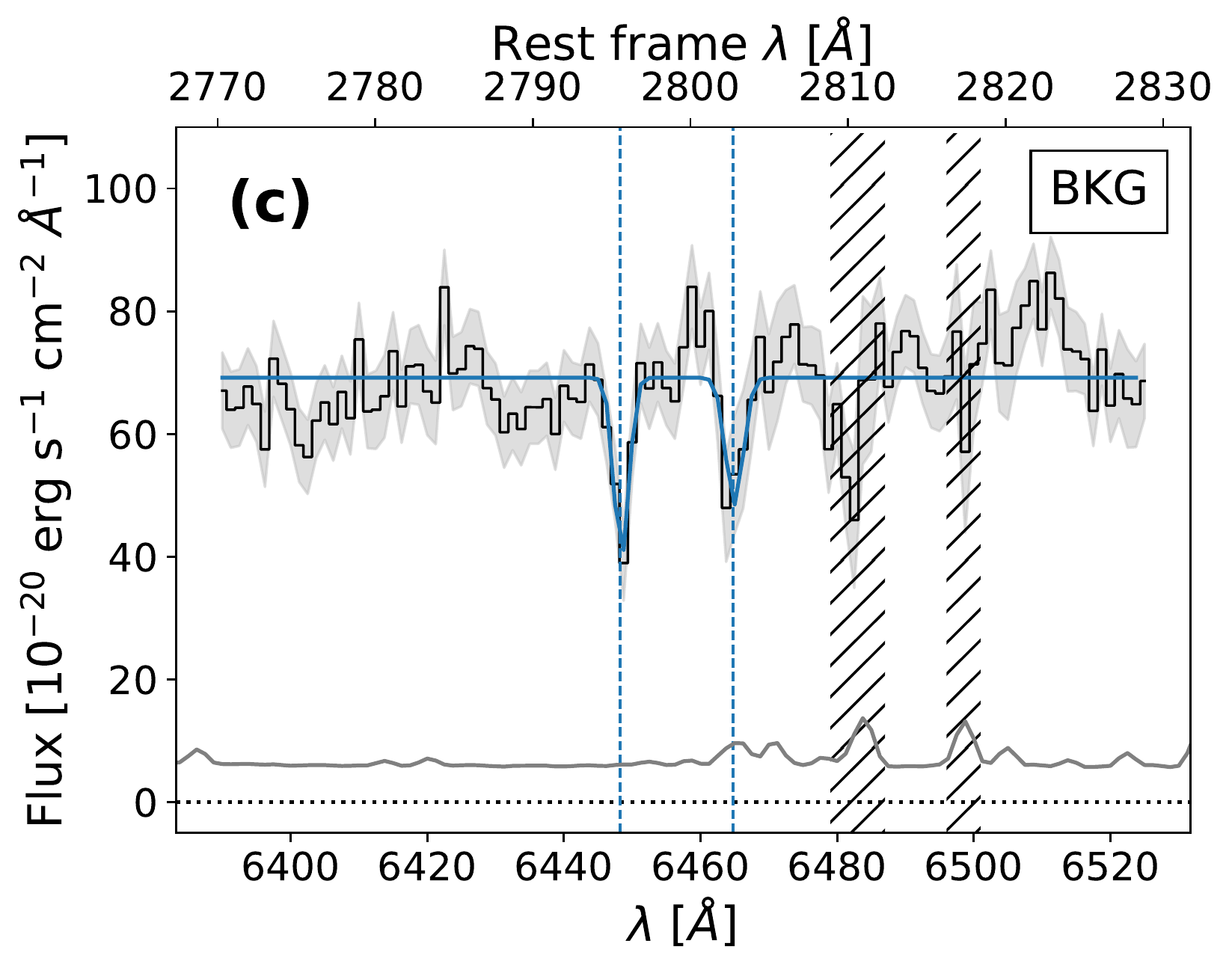}
    \end{subfigure} 
\end{minipage}
\caption{\textit{(a)} \mgii\ narrow-band image of the group (see Sect.~\ref{sec:41}) smoothed with a 0.6$\arcsec$ FWHM Gaussian and plotted with a power-law stretch. The white contours correspond to \mgii\ significance levels of 1.5, 2 and 3 $\sigma$ (solid, dashed and dotted, respectively) where 1$\sigma$ corresponds to a SB level of 1 $\times$ 10$^{-19}$ $\sbl$. The blue contours indicate the negative SB values (same coding as the positive contours) corresponding to the absorbing regions (see Sect.~\ref{sec:42}). 
The green contours trace the continuum of the group members close to \mgii\ and the neighboring galaxies as detected in the HST/ACS F775W image. The dotted green contour corresponds to the outer HST continuum contours of the group members put at MUSE resolution.
The FWHM of the MUSE PSF and smoothing kernel are shown with a hatched and empty circle, respectively, at the bottom right of the figure. 
\textit{(b)} Exposure map centered on the group. The solid, dashed and dotted black contours indicate 10, 30 and 90 hour depth, respectively. The white contour corresponds to the \mgii\ significance level of 1.5$\sigma$.
\textit{(c)} \mgii\ absorption in the spectrum of the background galaxy BKG located at an impact parameter of 19 kpc from galaxy D in projection and at the nebula redshift. 
The blue line is the best-fit model of the \mgii\ doublet absorption (Sect.~\ref{sec:42}).
The 1$\sigma$ uncertainties are shown in grey and the hatched areas indicate the position of strong sky lines and noise peaks.
The systemic redshift of galaxy D is indicated by vertical dashed lines.}
\label{fig:nb_mgii}
\end{figure*}

The NB image shown in Fig.~\ref{fig:nb_mgii}a has been smoothed using a 0\farcs6 (3 MUSE spaxels) FWHM Gaussian in order to increase the S/N while keeping the spatial resolution as good as possible (PSF of 0.52$\arcsec$ FWHM, see Fig.~\ref{fig:snrmap_smo} in Appendix~\ref{ap:1}). The 1$\sigma$ significance level corresponds to a surface brightness (SB) of 1 $\times$ 10$^{-19}$ $\sbl$. As shown in Fig.~\ref{fig:nb_mgii}b, the exposure time ranges from $\approx$ 30 to 90 hours through the \mgii\ nebula (white contour). This explains the noisier regions in the upper right corner of the NB image. As a consequence, we refrain from estimating the noise from empty regions around the nebula to compute the limiting SB contours and use the propagated variance of the MUSE cube in each pixel after smoothing. The total \mgii\ flux within 2$\sigma$ significance level is 7.48$\pm$0.50 $\times$ 10$^{-18}$ $\flcgs$ and corresponds to a total luminosity of 8.0$\pm$0.5 $\times$ 10$^{40}$ erg s$^{-1}$ without dust correction.

The eight 1.25 $\AA$ slices (from $-$177 to +286 \kms) of the MUSE subcube summed to create the $\lambda$2796 \mgii\ NB image are shown in Fig.~\ref{fig:NBlayer_mgii}. We note that because of the line spread function, those slice images are somewhat correlated. Despite that, spatial variations are still visible between adjacent slices (see Sect.~\ref{sec:44}).

\subsection{\mgii\ morphology}
\label{sec:42}

Figure~\ref{fig:nb_mgii}a shows the extended nature of the \mgii\ emission around this group of five galaxies. The observed nebula encompasses a projected area of 15 arcsec$^2$, corresponding to 1000 kpc$^{2}$ at the redshift of the group, above the 2$\sigma$ SB threshold of 2 $\times$ 10$^{-19}$ \sbl. The detected nebula resides within the virial radius of galaxy A (Sect.~\ref{sec:31}) and has an elongated shape linking the five group members with maximal projected extent reaching $\approx$70 kpc along the A/E direction and $\approx$40 kpc along the B/C direction. 

The strongest \mgii\ emission coincides spatially with galaxy C. Another emission peak is observed in between galaxies A and B. While globally the nebula appears aligned with the minor axis of galaxie(s) A (and E) as expected e.g. in a scenario where \mgii\ is emitted by biconical outflows launched perpendicular to the galaxy disk \citep{Bou12}, here the two intensity peaks are aligned with the major axis of the most massive galaxy A. This point is discussed in Sect.~\ref{sec:62}.

\begin{figure*}
\centering
   \resizebox{\hsize}{!}{\includegraphics{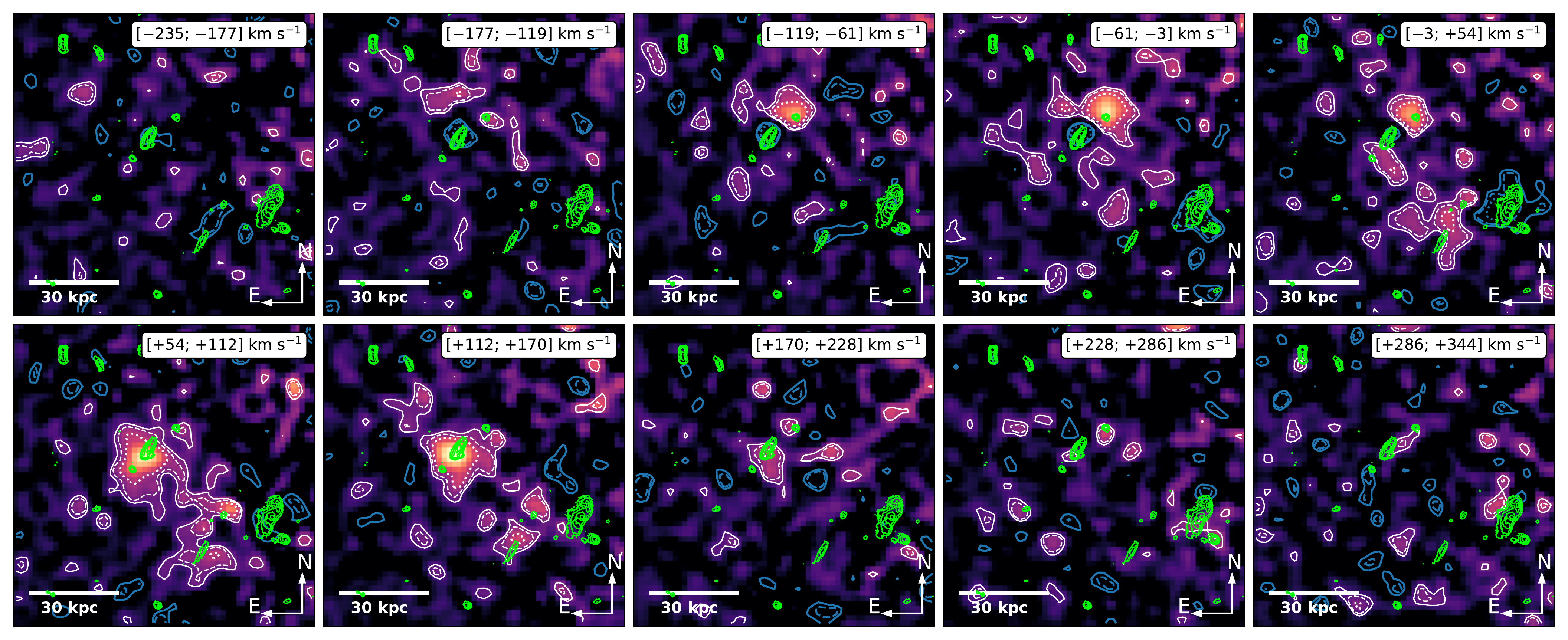}}
    \caption{Channel maps around the \mgii\ $\lambda$2796 emission. Each map corresponds to one MUSE subcube slice (1.25$\AA$ or 58 km s$^{-1}$ at the redshift of the group).
    The velocity window relative to the systemic redshift of galaxy A is indicated on each panel. Contours are the same as in Fig.~\ref{fig:nb_mgii}a. 
    The eight central slices (from $-177$ to $+286$ \kms) were summed to create the optimized $\lambda$2796 \mgii\ NB image shown in Fig.~\ref{fig:nb_mgii}a (see Sect.~\ref{sec:41}). }
    \label{fig:NBlayer_mgii}
\end{figure*}

The 2D map reveals a lack of flux at the positions of galaxies A and BKG (background galaxy). Those features are artificially created by the continuum subtraction procedure (see Sect.~\ref{sec:41}): the absorption features appear negative once the continuum has been subtracted.
The fact that \mgii\ in emission is observed at the position of galaxy A in the [+54; +112] and [+112; +170] \kms\ channel maps of \mgii~$\lambda2796$ (Fig.~\ref{fig:NBlayer_mgii}) indicates that both absorption and emission coexist in this region. The lack of flux detected at the position of galaxy A in the summed NB image (Fig.~\ref{fig:nb_mgii}a) therefore corresponds to regions where the absorption is dominant over the emission. As indicated in Sect~\ref{sec:33}, we do not aim at correcting for the emission infill effects here because our study focuses on the larger scale \mgii\ nebula.

Interestingly, while \mgii\ emission lines are detected in the integrated spectra of the galaxies D and E, the \mgii\ map shows no flux at the location of galaxy E and offset emission near galaxy D. This can be caused by the galaxy's absorption being completely filled by the emission (e.g. \citealt{Zabl21}). The spectra of the galaxies D and E indeed show hints of low S/N absorption lines which are within the \mgii\ NB spectral bandwidth (see Fig.~\ref{fig:integ_sp}). We also note that the segmentation map of galaxy D extends beyond the stellar body of the galaxy (Fig.~\ref{fig:integ_sp}). This implies that the \mgii\ emission observed in the spectrum of galaxy D likely originates from circumgalactic regions (as seen in the \mgii\ map in Fig.~\ref{fig:nb_mgii}a).

The \textsf{ODHIN} spectrum of BKG (z$\simeq$1.8463, see Fig.~\ref{fig:group_presentation}c) which is located outside the detected \mgii\ emission nebula and at a projected distance of 19 kpc from galaxy D, shows \mgii\ absorption at the same redshift as the nebula (see Fig.~\ref{fig:nb_mgii}c). 
Such a detection indicates that the \mgii\ gas even extends beyond the detected \mgii\ emission nebula. 
Using the line fitting procedure described in Sect.~\ref{sec:33}, we measured an \mgii\ $\lambda$2796 rest frame EW of 0.5$^{+0.6}_{-0.4}$ \AA\ for the absorption. 
This measurement falls well on the \mgii\ EW versus impact parameter relation found in several absorption studies (e.g. \citealt{Lundgren21}). 
Using the \cite{Menard09} relation between the \hi\ column density and the rest frame \mgii\ $\lambda$2796 EW we obtained the rough estimate of 10$^{19}$ atoms cm$^{-2}$ for the extended gas column density in this line of sight.
We refrain from analysing the absorption further because of the low S/N of the absorption lines.

The spatially resolved MUSE view of the nebula also reveals that the \mgii\ distribution is different from that of the continuum: the \mgii\ gas is more extended than the stellar content of the galaxies (see Sect.~\ref{sec:43}) and a projected bridge seems to link galaxy subgroups [A,B,C] and [D,E] (see Sect.~\ref{sec:43}). 
This bridge might not be a projection effect because of its tentative detection (S/N$\lesssim$2) in the single 1.25$\AA$ (58 km s$^{-1}$) cube slice between $+54$ and $+112$ km s$^{-1}$ relative to the systemic redshift of galaxy A (bottom left panel of Fig.~\ref{fig:NBlayer_mgii}).
We can also mention that the \mgii\ spatial distribution is not homogeneous in between the galaxies: several intensity peaks are detected at S/N>3. They suggest the presence of gas clumps or satellite galaxies (undetected in the HST and MUSE data) residing in the IGrM (see discussion in Sect.~\ref{sec:621}).

\subsection{Radial \mgii\ surface brightness profile}
\label{sec:43}

\begin{figure*}
\centering
   \resizebox{\hsize}{!}{\includegraphics{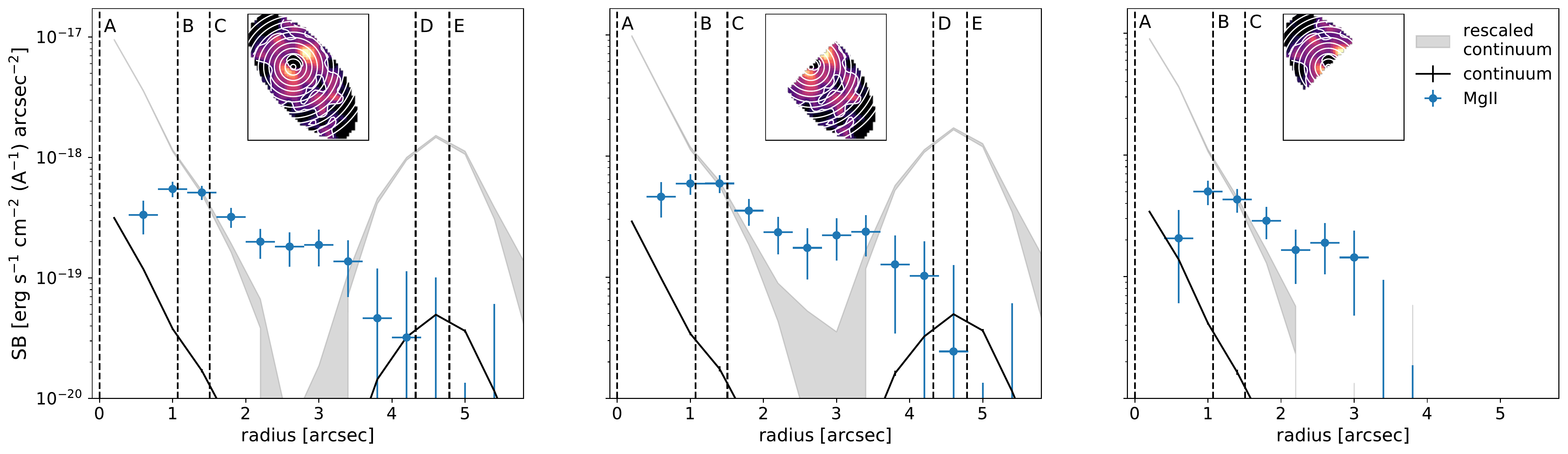}}
    \caption{Azimuthally averaged radial surface brightness profile of the \mgii\ nebula centered on galaxy A (blue data points). For comparison the profile of the \mgii\ continuum ($\pm$20 000 $\kms$ around the \mgii\ doublet) shown in black is also rescaled (grey) to the \mgii\ SB at the location of galaxy C (where \mgii\ absorption is unlikely, see Appendix~\ref{ap:3}). The positions of the group members are indicated with vertical dashed lines. The radial profiles have been calculated using unsmoothed \mgii\ and continuum images after masking the pixels outside the 1$\sigma$ significance level approximated with an elliptical contour as shown in the inset (Left). The middle and right panels show the radial profiles after masking the pixels above and below the [A,B,C] axis, respectively.
    In the insets, we also show the 2-pixel annuli used to construct the profiles as white circles.
    }
    \label{fig:sbprof_mgii}
\end{figure*}

In order to highlight the extended nature of the \mgii\ emission around and in between the galaxy group members, we computed azimuthally averaged radial surface brightness profiles of the \mgii\ and continuum emission centered on galaxy A (Fig.~\ref{fig:sbprof_mgii}, first panel). The \mgii\ profile (blue) is measured on the unsmoothed \mgii\ NB image (Sect.~\ref{sec:41}) after masking pixels located outside an ellipse corresponding to the 1$\sigma$ significance level in order to increase the S/N in the two-pixel-wide annuli (or truncated annuli) used for the aperture photometry (see inset panels). The width of the annuli (0.4") is comparable to the spatial resolution element of the data (PSF FWHM of 0.52").

The \mgii\ continuum radial SB profile (black) is measured from a continuum image extracted from the continuum-only cube within a window of $\pm$20 000 $\kms$ around the \mgii\ doublet. This broad spectral window does not include strong absorption or emission lines and ensures a good enough S/N. The neighboring sources were rigorously masked in order to avoid contamination and the same elliptical mask used on the \mgii\ image was applied to compute the continuum SB profile. Errors were measured in each (truncated) annulus using the estimated variance from the MUSE data cube.
To aid the visual comparison, the continuum profile has been rescaled to the \mgii\ SB at the position of galaxy C which is where the brightest \mgii\ intensity peak is observed. Galaxy C is unlikely to be affected by \mgii\ absorption as its measured \mgii\ line ratio $\lambda$2796$\AA$/$\lambda$2803$\AA$ is very close to two, which is the expected value for optically thin emission (see Appendix~\ref{ap:3}). Moreover, no fluorescent \feii\ lines are significantly detected for galaxy C which is in good agreement with no infilling effect and therefore no absorption (see Appendix~\ref{ap:3}, \citealt{Mauerhofer21}).

The \mgii\ SB profile appears more extended than the continuum, especially in between the galaxies [A,B,C] and [D,E] -- radial positions are indicated by vertical dashed lines -- where a low SB projected "bridge" at $\approx$2$\times$10$^{-19}$ $\sbl$ level can be identified. 
In order to differentiate between the extended emission around the galaxy subgroup [A,B,C] and in between the galaxy subgroups [A,B,C] and [D,E], we compute the radial SB profiles above and below the [A,B,C] axis (see insets for illustration) in the middle and right panels of Fig.~\ref{fig:sbprof_mgii}, respectively.
The \mgii\ emission appears more extended than the rescaled continuum in both directions (above and below the [A,B,C] axis), confirming the detection of diffuse \mgii\ emission around the group members as well as a low SB \mgii\ bridge ($\approx$2$\times$10$^{-19}$ $\sbl$) detected with more than 1.5$\sigma$ significance spanning $\approx$50 kpc in projection between the galaxy subgroups [A,B,C] and [D,E] (see Fig.~\ref{fig:nb_mgii}a).

We go further into the analysis of the \mgii\ nebula in the next section by looking at the spatial variation of the \mgii\ spectral properties.

\subsection{Spatially resolved \mgii\ properties}
\label{sec:44}

The kinematics of the gas connecting the galaxy group can provide crucial information to understand its nature and origin. 

In order to reveal the spatial variations of the \mgii\ spectral profile within the nebula, we build a 2D binned map using the weighted Voronoi tesselation method \citep{C03,Diehl&Statler06}. 
This method allows us to increase the S/N in the surroundings of the galaxies where the surface brightness is the lowest. The binning is performed on the \mgii\ NB image constructed from the MXDF datacube (see Sect.~\ref{sec:41}) for which the pixels with S/N < 1.5 (corresponding to the outer SB contour in Fig.~\ref{fig:nb_mgii}a) have been masked in order to remove the noise-dominated pixels. The resulting binned \mgii\ map consists of 20 bins with S/N > 3 and probes the detected \mgii\ nebula above $\approx$ 2 $\times$ 10$^{-19}$ \sbl.

The spectral extraction is performed from the MUSE continuum-subtracted cube smoothed with a 0.6\arcsec\ FWHM Gaussian in order to increase the S/N while keeping a good spatial resolution (see Appendix~\ref{ap:1}).
We extract the \mgii\ doublet in each resulting bin and measure the properties of the emission lines by modelling the doublet as a sum of two Gaussian profiles with fixed peak separation.

The errors on the line parameters (peak position and FWHM) were estimated using bootstrapping: for each segmented area, we generated 1000 realizations of the extracted line where each pixel is randomly drawn from a normal distribution centered on the original pixel value and with standard deviation derived from the estimated noise value. The noise of each extracted spectrum corresponds to the standard deviation of the data measured in two 200 $\AA$ wide spectral windows located on each side of the \mgii\ doublet. 
We note that because of smoothing and because of the PSF the line parameters between neighboring pixels are somewhat correlated. However, spatial variations are still visible.

\begin{figure*}
\begin{minipage}{0.29\textwidth}\centering
   \resizebox{\hsize}{!}{\includegraphics{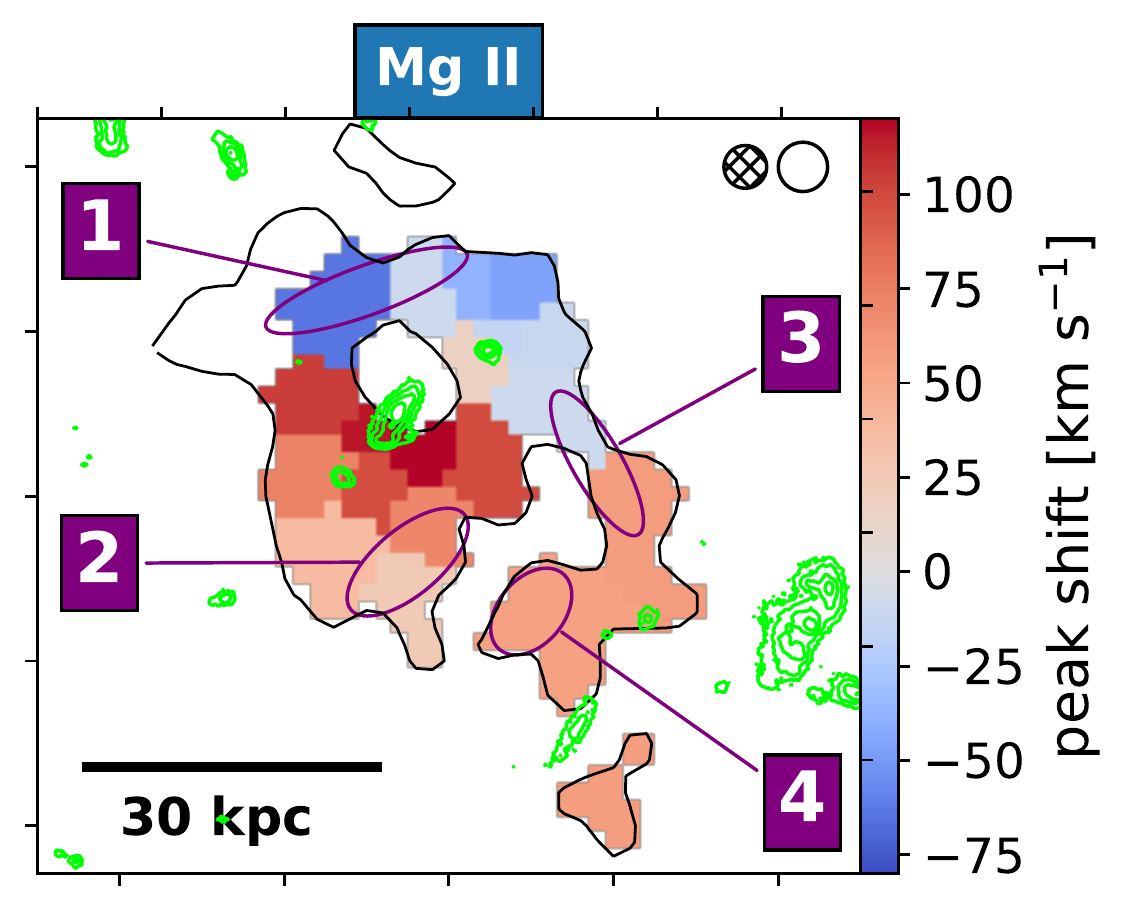}}
\end{minipage}
\begin{minipage}{0.71\textwidth}
 \centering
   \includegraphics[width=1\textwidth]{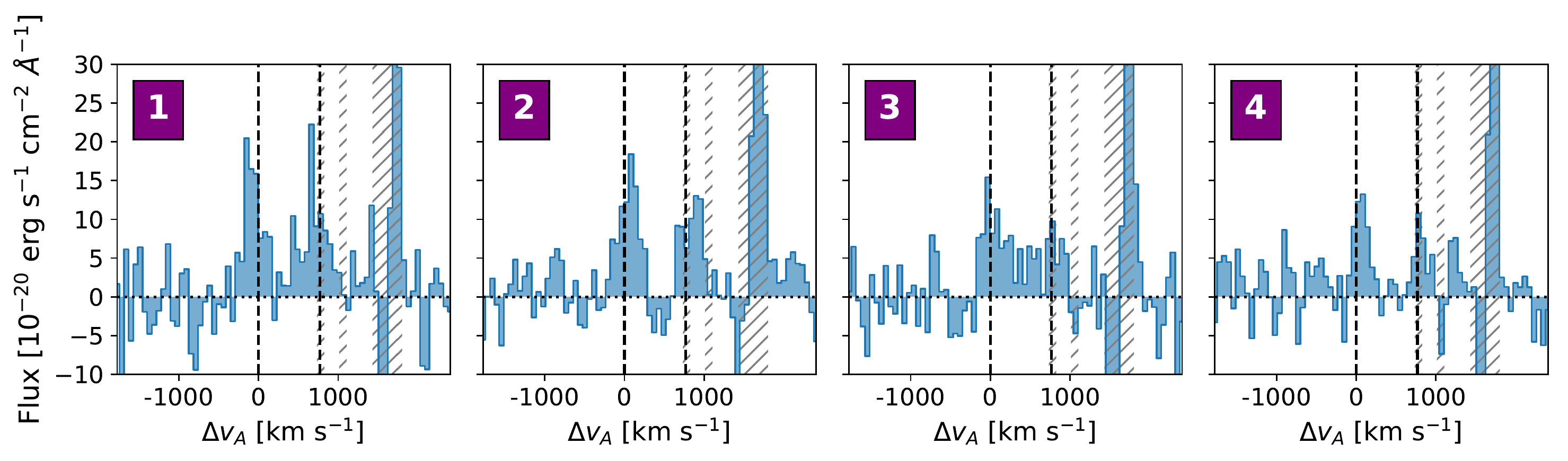}
\end{minipage}
\begin{minipage}{0.29\textwidth}\centering
   \resizebox{\hsize}{!}{\includegraphics{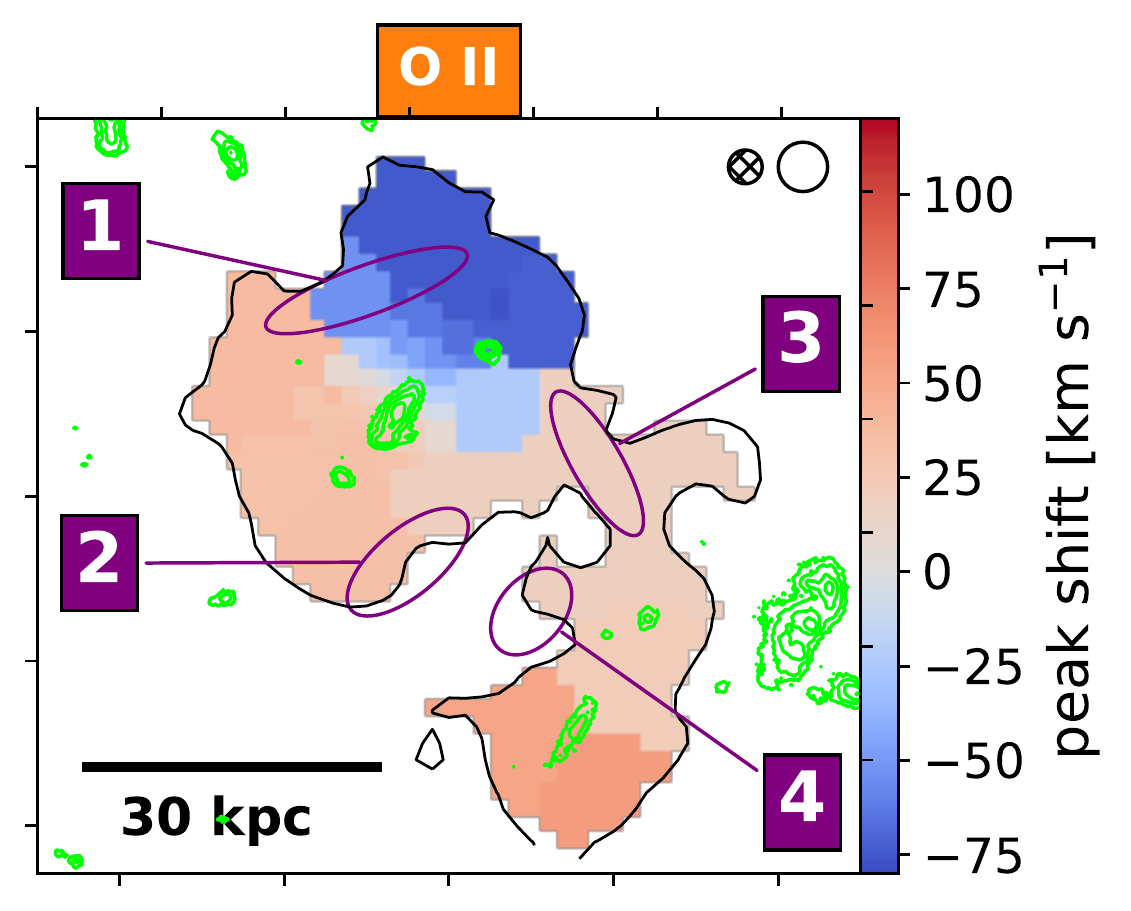}}
\end{minipage}
\begin{minipage}{0.71\textwidth}
 \centering
   \includegraphics[width=1\textwidth]{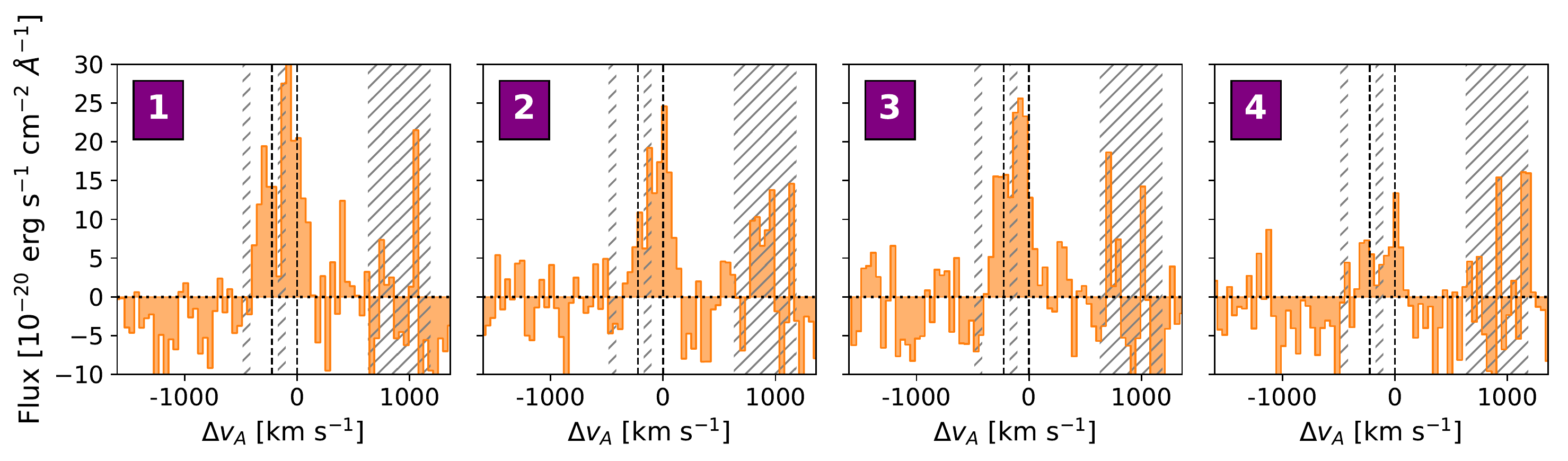}
\end{minipage}
\caption{\textit{Left:} binned line-of-sight velocity maps of the \mgii\ (top) and \oii\ (bottom) emission relative to the systemic redshift of galaxy A (Sects.~\ref{sec:44} and \ref{sec:51}). The black contours correspond to the 1.5$\sigma$ significance level. The green contours trace the galaxies detected in the HST/ACS F775W image. The FWHM of the MUSE PSF and smoothing kernel are shown with a hatched and an empty circle, respectively, at the top right of the panels. 
The diverging \mgii\ and \oii\ colormaps are centered on the systemic redshift of galaxy A and have the same dynamical range to ease the visual comparison (Sect.~\ref{sec:51}). 
\textit{Right:} \mgii\ (top) and \oii\ (bottom) doublet extracted in the elliptical areas traced in purple and designated by the same number on the map (left panels). The vertical dashed lines correspond to the systemic redshift of galaxy A. The vertical hatched bands indicate the presence of sky lines.}
\label{fig:velmap_mgii}
\end{figure*}

Figure~\ref{fig:velmap_mgii} (top left panel) shows the resolved map of the \mgii\ $\lambda$2796 peak position relative to the systemic redshift of galaxy A. 
The observed \mgii\ velocity gradient suggests an overall rotation of the structure along the major axis of galaxy A consistent with an extension of the ISM rotation (as probed by the \oii\ emission, see Sect.~\ref{sec:51}) at the CGM scale within which the satellite galaxies B and C are embedded. This ISM/CGM co-rotation is also observed in simulations and observations (e.g. \citealt{S19,Zabl19,Ho20,Zabl21}). 

The top right panels of Fig.~\ref{fig:velmap_mgii} show the unsmoothed \mgii\ doublet extracted in the elliptical areas traced in purple and designated by the same number on the map. These apertures are independent of the Voronoi bins and aim to show the \mgii\ line profile in some diffuse regions of the nebula. 
The velocity variations are small (< 50 $\kms$) in the diffuse regions. 

The kinematics of the nebula appear to be overall shifted towards higher velocities (by $+$50 km s$^{-1}$ on average) compared to the \oii\ emission (bottom left panel, see Sect.~\ref{sec:51}). 
The highest velocity variations are observed along the minor axis of galaxy A and reach $\approx$+120 $\kms$ and $-$80 \kms with respect to the systemic redshift of galaxy A. As discussed in Sect.~\ref{sec:621}), this supports the presence of an outflow emerging from galaxy A.

\section{\oii\ and \feii\ emission}
\label{sec:5}

At the redshift of the group, MUSE covers the non-resonant collisionally excited \oii\ doublet and the fluorescent \feii\ lines. The corresponding NB images and radial SB profiles are shown in Fig.~\ref{fig:oiifeii_nbsb} and are created following the same procedure as for the \mgii\ emission (Sects.~\ref{sec:41} and \ref{sec:43}, respectively).

\subsection{A low surface brightness \oii\ nebula}
\label{sec:51}

\begin{figure*}
\centering
   \resizebox{0.94\hsize}{!}{\includegraphics{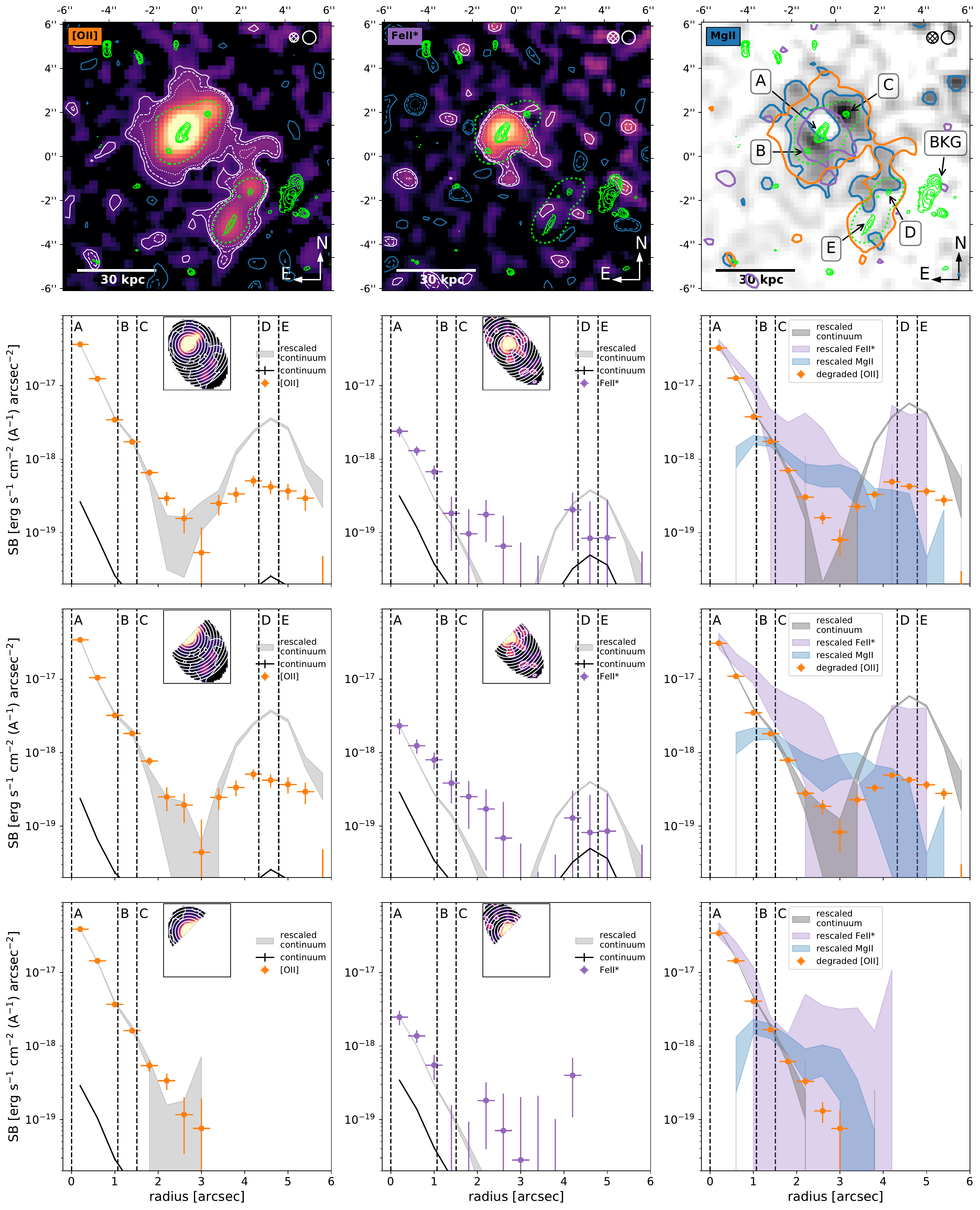}}
    \caption{\textit{First column:} \oii\ NB image (top, Sect.~\ref{sec:51}) with legend being the same as in Fig.~\ref{fig:nb_mgii}a. The 1$\sigma$ level corresponds to a SB of 2.5 $\times$ 10$^{-19}$ \sbl. The bottom panels show the total and directional (below and above the [A,B,C] axis) azimuthally averaged radial SB profiles centered on galaxy A (from top to bottom, respectively, see also insets for illustration) of the \oii, \oii\ continuum and rescaled \oii\ continuum (orange, black and grey, respectively) emission. The radial positions of other group members with respect to galaxy A are indicated by the vertical dashed lines. 
    \textit{Second column:} Same as first column but for \feii\ (purple). 
    \textit{Third column:} Comparison of the \mgii, \oii\ and \feii\ 2$\sigma$ significance level contours (blue, orange and purple, respectively) superimposed on the \mgii\ NB image (top). The bottom panels show a comparison of the radial SB profiles where the \feii\ profile (shaded purple) has been rescaled to the \oii\ profile at the SB peak and \mgii\ (shaded blue) at the position of galaxy C (Appendix~\ref{ap:3}). For this plot the \oii\ map was degraded to the resolution of the \mgii\ emission (Sect.~\ref{sec:51}).}
    \label{fig:oiifeii_nbsb}
\end{figure*}

The \oii\ NB map (Fig.~\ref{fig:oiifeii_nbsb}, top left) is obtained by summing the continuum-subtracted cube in the wavelength range [8590, 8605] $\AA$ ($\approx$ 500 \kms). In order to improve the S/N, the image was smoothed using a 0\farcs6 FWHM Gaussian function. The white contours correspond to the 1.5, 2, and 3 $\sigma$ significance levels with 1$\sigma$ corresponding to 2.5$\times10^{-19}$ $\sbl$. The total \oii\ flux and luminosity within the 2$\sigma$ significance level contour (dashed white line) are 5.95$\pm$0.07$\times10^{-17}$ \flcgs\ and 6.36$\pm$0.07$\times10^{41}$ erg s$^{-1}$, respectively.

The \oii\ azimuthally averaged radial SB profiles are shown in the lower left panels of Fig.~\ref{fig:oiifeii_nbsb}. The profiles are constructed following the same procedure as for \mgii\ (see Sect.~\ref{sec:43}) with an additional step for the construction of the continuum image where the cube slices within $\pm$20 000 km s$^{-1}$ polluted by skylines have been masked. This is particularly important because at the redshift of the nebula, the \oii\ line falls next to many sky lines. 
The \oii\ doublet is itself contaminated by a skyline (see vertical grey line in the lower right panels of Fig.~\ref{fig:velmap_mgii}) preventing us from in-depth analysis.

When comparing the \oii\ and rescaled continuum radial SB profiles in Fig.~\ref{fig:oiifeii_nbsb}, only one data point is above the rescaled continuum profile with more than 2$\sigma$ significance. The inner profile has the same shape as the continuum one implying that the \oii\ emission is not significantly more extended than the continuum inside a radius of $\approx$ 2\arcsec\ from galaxy A.
Extended \oii\ emission at the same position as the \mgii\ bridge, i.e. in between the galaxies [A,B,C] and [D,E], is detected with 2$\sigma$ significance level in the \oii\ NB image (Fig.~\ref{fig:oiifeii_nbsb}, top left). However, due to averaging effects, the \oii\ bridge is less significant on the radial profiles computed below the [A,B,C] axis, i.e. in the direction of the [D,E] subgroup (first column third row of Fig.~\ref{fig:oiifeii_nbsb}).
The radial SB profile computed above the galaxies A, B and C (bottom left panel) indicates extended \oii\ emission at radii > 2.5\arcsec\ from galaxy A, suggesting the presence of an ionized \oii\ halo around this galaxy subgroup.
Comparisons of the \mgii\ and \oii\ spatial extent and radial SB profiles are presented in Fig.~\ref{fig:oiifeii_nbsb} (top two right panels). Since the MUSE PSF is narrower at the \oii\ wavelength than at the \mgii\ wavelength, we degraded the \oii\ NB image by convolving it by a Moffat function whose parameters ($a=0.22$ and $b=2.87$, see \citealt{B17}) have been empirically determined so that the \oii\ and \mgii\ images have the same PSF. While the \oii\ and \mgii\ spatial distributions are consistent, the \mgii\ radial SB profiles (azimuthally averaged and in directions, bottom right panels) are all flatter than the \oii\ profiles. This can be attributed to the resonant nature of the \mgii\ transition (see Sect.~\ref{sec:622}).

The binned line-of-sight velocity map of the \oii\ emission is shown in the bottom left panel of Fig.~\ref{fig:velmap_mgii}. The map is constructed following the same procedure as for \mgii\ (Sect.~\ref{sec:44}) and consists of more bins (57) because of the higher S/N of the \oii\ emission near the galaxies. 
At the position of the galaxies, the observed spatial variations of the \oii\ peak positions are consistent with their redshift relative to galaxy A (right panel of Fig.~\ref{fig:group_presentation}).
The lower right panels of Fig.~\ref{fig:velmap_mgii} show the unsmoothed \oii\ lines extracted in the same elliptical apertures as for \mgii\ (top panels). Interestingly, in region \#4 the \mgii\ emission is stronger than \oii. This can be explained by the resonant nature of the \mgii\ transition (see Sect.~\ref{sec:622}) resulting in \mgii\ being more extended that \oii.
The existence of the \oii\ bridge is reinforced by the detection of the \oii\ doublet in this area (region \#3 in Fig.~\ref{fig:velmap_mgii}).

\subsection{Extended \feii\ emission}
\label{sec:52}

The \feii\ NB image (top middle panel of Fig.~\ref{fig:oiifeii_nbsb}) is obtained by summing the $\lambda$2365, $\lambda$2396, $\lambda$2612 and $\lambda$2626 NB images, of which the observed spectral windows are [5448.75, 5456.25], [5522.5, 5530.0], [6021.25, 6027.5] and [6041.25, 6066.25] $\AA$, respectively. The total \feii\ flux and luminosity within the significance level contour of 2$\sigma$ is 5.3$\pm0.5\times10^{-18}$ \flcgs\ and 5.7$\pm$0.6$\times10^{40}$ erg s$^{-1}$, respectively. 

Given that the \mgii\ and \feii\ lines are close in wavelength, we consider the same stellar continuum (see Sect.~\ref{sec:41}).
Therefore, the radial SB profiles of the continuum (black profile in the middle bottom panels) are the same as for \mgii. The comparison between the \feii\ emission and the rescaled continuum radial SB profile reveals that the \feii\ emission is more extended than the continuum (see central panel of Fig.~\ref{fig:oiifeii_nbsb}) second row. The \feii\ extended emission appears more significant in the direction of the galaxies D and E (compare the central panels of the Fig.~\ref{fig:oiifeii_nbsb} third and fourth rows).
This is confirmed in the NB image where the \feii\ emission indeed extends along the minor axis of galaxy A and towards galaxy D. We discuss the origin for the extended nature of the \feii\ emission in Sect.~\ref{sec:62}. Contrary to \mgii\ and \oii\ emission, the \feii\ emission does not surround the five-galaxy group but is rather centered on the most massive galaxy A (see the \mgii, \oii\ and \feii\ contours and radial SB profiles comparison in the last column of Fig.~\ref{fig:oiifeii_nbsb}). 

Similarly, \cite{Finley17} using MUSE found the \feii\ emission of a $z=1.29$ galaxy to be more extended than the stellar continuum and aligned with the minor axis of the galaxy. The velocity gradient and shape of the \feii\ emission suggest the presence of a conical outflow. The faintness of the \feii\ emission of galaxy A hampers a kinematic analysis.

\section{Discussion}
\label{sec:6}

\subsection{Existence of an Mg-enriched intragroup medium}
\label{sec:61}

Historically the existence of an intragroup medium enriched in magnesium has been suggested by absorption studies in order to explain the fact that, while the strong \mgii\ absorbers (rest frame equivalent width EW>0.8$\AA$) are generally associated to a single galaxy (within 100 kpc, e.g. \citealt{Bouche07, Bouche12, Ho17, Schroetter16, Schroetter19, Zabl19, Lundgren21}), sometimes weaker \mgii\ absorbers (EW < 1$\AA$, e.g. \citealt{Muzahid18,Du20}) or \hi\ absorbers (e.g. \citealt{P17,Ra18}) can be matched to groups of multiple galaxies ($\Delta$r < 500 kpc and $\Delta$v < 500 $\kms$), in agreement with clustering studies \citep{Bouche06,Lundgren09,Gauthier09}.

Whether observed absorption features (with EW < 1$\AA$) imprinted in the spectrum of a background source are tracing gas coupled to the group or circumgalactic gas of an individual galaxy is still a matter of debate. 
Indeed, absorption studies reported contradictory results, some suggesting that \mgii\ absorbing systems were associated with the CGM of the individual group members (e.g. \citealt{B11,Fo19}) and others to a widespread IGrM (e.g. \citealt{Bi17, P17, N18}) originating from a mixture of previous tidal interactions between group members and outflowing winds. Disentangling the two is very challenging, especially when there is only one-dimensional information for a given system.

Our discovery of the first extended \mgii\ nebula surrounding and connecting the members of a 5-galaxy group provides a panoramic view of the neutral enriched gas residing in between galaxies. This group is rather small and compact (see Sect.~\ref{sec:31}), meaning that it is likely an interacting system where the CGM of individual galaxies are mixed. This idea is reinforced by the fact that the detected \mgii\ nebula resides inside the virial radius of galaxy A estimated at $\approx$100 kpc.
Moreover, the detection of a gaseous bridge in between the two subgroups [A,B,C] and [D,E] suggests that we are observing a widespread low surface brightness structure embracing the five galaxies, also called \textit{intragroup} medium. We note that there is no precise definition of the IGrM in the literature. 
Our analysis brings the first observational evidence for a low SB diffuse component of neutral gas residing within a galaxy group at $z>1$ that we define as the intragroup medium. 

The detection of \mgii\ absorption in the spectrum of the background source BKG (see Sect.~\ref{sec:42}) indicates that the IGrM is actually even more extended than the detected nebula. Our EW measurement of the \mgii\ absorption (Sect.~\ref{sec:42}) is in good agreement with the EW versus impact parameter relation established by quasar absorption line studies (e.g. \citealt{Lundgren21}). This result highlights the complementarity of the absorption and emission methods to study the gas around galaxies.

\subsection{Origin of the intragroup nebula}
\label{sec:62}

In this section we discuss two points: the presence of enriched gas beyond the galaxies and the mechanisms that make it shine.

\subsubsection{Why is there enriched gas so far from the stars ?}
\label{sec:621}

Several scenarios have been suggested to explain the presence of magnesium-enriched gas in between the galaxies of a group. Magnesium atoms are $\alpha$ elements released in the ISM and CGM by core collapse supernovae. One can therefore naturally imagine that the magnesium gas residing outside galaxies has been ejected through strong supernovae-driven outflows. The P-Cygni shape of the \mgii\ doublet in the integrated spectrum of Galaxy A (see Fig.~\ref{fig:integ_sp} top) is in good agreement with this scenario. Indeed, according to radiative transfer (RT) models, this line profile is a signature of resonant scattering in an optically thick outflowing medium where the redshifted emission corresponds to back-scattered photons reflected by the receding medium \citep{H00,D06,V06,K10}. The \mgii\ lines of galaxies D and E seem to be redshifted compared to their systemic redshifts (Fig.~\ref{fig:integ_sp}), however the continuum is too faint to detect significant absorption lines and therefore P-Cygni profiles, i.e. the presence of outflows for those galaxies.
We detect >2$\sigma$ \mgii\ emission aligned with the minor axis of galaxy E and near galaxy D (Fig.~\ref{fig:nb_mgii}). These regions could have been enriched via outflows after supernovae feedback. However, the S/N of our data is too poor in these areas to carry out a resolved kinematics study.
The resolved \mgii\ map (Fig.~\ref{fig:velmap_mgii}) shows that the two regions along the minor axis of galaxy A have different velocities with opposite signs ($\approx$ $+120$ and $-$80 \kms\ with respect to the systemic redshift of galaxy A), providing additional evidence for the presence of an outflow. 
Even if we do not find clear evidence for an outflow in the four other galaxies, we cannot exclude the deposition of metals in the CGM by past feedback processes explaining the observed diffuse \mgii\ emission detected around each of the galaxy members. 
In that scenario, one can imagine that the low surface brightness gaseous bridge in-between the galaxy subgroups [A,B,C] and [D,E] could be due to galactic wind transfer between the group members as observed in the FIRE simulation \citep{A17}. 

The proximity of the galaxies, as well as the detection of a bridge in-between the galaxy subgroups is also compatible with another scenario where the magnesium gas has been tidally stripped out of galaxies due to past gravitational interactions among the galaxy group members. In this framework, the detected enriched bridge would correspond to a tidal feature linking the galaxies. 
In this scenario, \cite{Bi17} suggested that the IGrM could consist of multiple cool gas systems orbiting around galaxies to form the IGrM. As mentioned in Sect.~\ref{sec:42}, we detect several S/N>3 \mgii\ intensity peaks within the IGrM, which, if real, could be associated to the cool gas clouds described in \cite{Bi17}. According to hydrodynamical simulations, such cold clumps could originate from gas density perturbations after tidal interactions \citep{N20}.
As mentioned above, one could also argue that the gas in the bridge could be outflowing material from galaxy A undergoing an "intergalactic transfer" to galaxy D \citep{A17}. 
However, the fact that neither the fluorescent \feii\ or non-resonant \oii\ emission are symmetrically distributed around galaxy A but extend preferentially towards galaxy D and E (see top panels of Fig.~\ref{fig:oiifeii_nbsb}) suggests that the enriched gas detected in the bridge has been stripped out of galaxies by tidal or ram pressure forces.
Moreover, we do not observe strong velocity gradients in the bridge meaning that the bridge dynamics are consistent with the rest of the structure (left panel of Fig.~\ref{fig:velmap_mgii}) and reinforcing the presence of tidally disrupted material. It is however more difficult to disentangle the different scenarios in the close surroundings of the galaxies where -- given the very small physical separation between the sources -- outflows, tidal interactions and intergalactic transfer could play a role.

Given the proximity of the five group members, we suggest that past tidal interactions are likely the dominant mechanisms explaining the presence of magnesium-enriched gas in the IGrM. Plenty of cool stripped gas is actually observed in simulations in the form of tidal features (e.g. \citealt{RB12,N20}). In \cite{RB12}, the $\approx$10$^{11.5}$ M$_{\odot}$ DM halo (similar to our system) at $z=3$ indeed shows orbiting satellites and tidal tails.
Such \hi\ tidally disruptive material is also observable in the local Universe, e.g. in-between the galaxies of the M81 system \citep{Y94}.

\subsubsection{Which mechanism makes the intragroup magnesium gas shine ?}
\label{sec:622}

Several mechanisms can explain the emission of \mgii\ photons from the intragroup gas. 
The possible sources, which can arise at all scales (ISM, CGM, IGrM), are the following: 
\begin{itemize}
    \item the stellar continuum at $\lambda \sim 2800$\AA: by summing the continuum contribution of the five galaxies in the group, we estimate the \mgii\ photon budget from the stellar continuum to 9.0$\pm$1.0 $\times$ 10$^{-18}$ $\flcgs$ corresponding to a total luminosity of 9.6$\pm$1.1 $\times$ 10$^{40}$ erg s$^{-1}$ without dust correction and considering the sum of the two doublet lines (see Sect.~\ref{sec:41}).
    \item photo-ionisation by \mgii\ ionising radiation, $\lambda < 825$ \AA, from the stellar continuum of galaxies in the group or from the UV background (UVB) and collisional ionisation of Mg$^{+}$ ions into Mg$^{2+}$ ions, followed by recombination cascades leading to the emission of nebular \mgii\ photons. Using \textsf{Cloudy} photo-ionisation models, we estimate the nebular \mgii\ photon budget corresponding to collisions and photo-ionisation from the stellar continuum (see Appendix~\ref{ap:5}).
    \item \mgii\ photons can be produced by shocks resulting from gravitational interactions or galactic outflows from galaxies.
\end{itemize}

The \mgii\ doublet is a resonant line meaning that the \mgii\ photons are likely to be scattered within an Mg$^+$ gas cloud. This property is usually invoked to explain the diffuse resonant emission around galaxies, e.g. $\lya$ halos (e.g. \citealt{K19,L20}). However, this process erases the information on the location where the photons have been emitted. In the following paragraphs, we discuss the possible distributions of sources and mechanisms that can explain the intragroup \mgii\ luminosity, and light distribution.

\paragraph{\bf Central source with scattering --}

{\it Stellar continuum:} The first possibility to explain extended \mgii\ emission is the scattering of stellar continuum photons, i.e. the re-emission of continuum photons absorbed in the \mgii\ transition. While Wisotzki et al. in prep. found that this process is likely the dominant one in their system, \cite{Zabl21} found that it is, under the assumption of a biconical outflow, likely not enough to explain the brightness of the \mgii\ halo. In the group studied here, the \mgii\ continuum luminosity (9.6$\pm$1.1 $\times$ 10$^{40}$ erg s$^{-1}$ without dust correction and integrating over the same velocity range as the \mgii\ NB image, see Sect.~\ref{sec:41}) is enough to explain the observed \mgii\ nebula (total luminosity of 8.0$\pm$0.5 $\times$ 10$^{40}$ erg s$^{-1}$). This suggests that the continuum scattering scenario is an important process at play in this system.

{\it Nebular emission:} Diffuse \mgii\ emission can also be produced by the scattering of nebular \mgii\ photons produced in the ISM of galaxies. For this mechanism, the \mgii\ photons are produced by collisions and recombinations associated with stellar UV radiation in star-forming regions of galaxies. The escaping photons can then scatter into the surrounding Mg$^+$ gas and some are redirected towards the observer. 
We used \textsf{Cloudy} models \citep{Ferland2017} to estimate the intrinsic \mgii\ flux budget produced by the stars of the five galaxy group members. We found an intrinsic \mgii/\oii\ flux ratio varying between $\sim$10 and 30\% (see Appendix~\ref{ap:5} for more details). The total observed \mgii/\oii\ flux ratio of the nebula (12\%) being lower than the intrinsic one, or similar depending on the models, we conclude that there is enough nebular \mgii\ emission to explain the observed \mgii\ extended emission through resonant scattering of photons originally produced in \hii\ regions. This scenario is reinforced by the spectral shapes of the galaxies (i.e. P-Cygni and redshifted lines with respect to the systemic redshift), suggesting that \mgii\ photons scattered in an outflowing medium (see Sect.~\ref{sec:621}).

\paragraph{\bf In situ emission --}

{\it Collisions + photo-ionisation:} Another scenario to explain the extended \mgii\ emission is that enough \mgii\ ionizing photons ($\lambda_0$<825\AA) escape the ISM of galaxies -- perhaps facilitated by galaxy-scale outflows -- so that \mgii\ photons are produced through photo-ionization in situ, i.e. directly in the the CGM/IGrM. It is however well known that the escape of ionizing photons from galaxies is very low (e.g. \citealt{Malkan03,Bridge10,Rutkowski16,Alavi20}).
The spatial coincidence of the \mgii\ and the non-resonant collisionally excited \oii\ emission lines yet supports this "non-scattering" origin.
We note that the resulting \mgii\ photons can then scatter in the surrounding magnesium located in the IGrM, as in the first scenario. This can explain the "butterfly" shape of the ionised \mgii\ nebula and the fact that the \mgii\ emission has flatter radial SB profiles than the \oii\ profiles (right panels of Fig.~\ref{fig:oiifeii_nbsb} and Sect.~\ref{sec:51}).

{\it UVB:} The UVB photons emitted by all the galaxies and quasars of the Universe at the redshift of the group with energy higher than 15 eV (i.e. $\lambda_0$ < 825\AA) can also ionize Mg$^{+}$ in the IGrM and thus produce \mgii\ photons. We compare the contributions of the UVB at z=1.3 \citep{HaardtMadau12} and the total observed (i.e. dust attenuated) stellar emission of the group members in Fig.~\ref{ap:52} (see also Appendix~\ref{ap:5}). By considering a very low escape fraction of the Mg$^{+}$ ionizing flux (0.7\%) we find that the UVB dominates the Mg$^{+}$ ionizing budget in the IGrM (i.e. $r$ > 10 kpc from the stars). In the vicinity of the stars (r$\approx$10 kpc), the contribution of the stellar emission becomes dominant. Therefore the UVB appears to be a non negligible source of ionization at large distances from the stars, i.e. in the CGM/IGrM., if we assume no escape of \mgii\ ionizing photons from the galaxies. When considering that a non-zero fraction of the ionizing flux escapes (e.g. 14\% in Fig.~\ref{ap:52}), the contribution of the UVB to the powering of the \mgii\ nebula becomes negligible. While, as mentioned above, the ionizing escape fraction in galaxies is usually found to be very low, galaxy C shows some hints for a clear line of sight (see Appendix~\ref{ap:3}), thus indicating the possibility of ionizing flux leakage (e.g. \citealt{C20}). A precise measurement of the ionizing flux escape fraction from the group members is needed to go deeper in the analysis but is beyond the scope of this paper.

{\it Satellite:} In order to explain the \mgii\ emission at the position of the bridge, we estimated upper limits of $\approx$10$^7$ M$_\odot$ and $\approx$10$^{-3}$ M$_\odot$ yr$^{-1}$ for the stellar mass and SFR, respectively \citep{Madau98,W14}, of an undetected galaxy at $z\approx1.3$ in the extremely deep F775W HST image (limiting magnitude of 29.5 mag corresponding to a UV luminosity of 3$\times$10$^{37}$ erg s$^{-1}$, \citealt{R15}). 
In other words, if there are unseen galaxies powering the \mgii\ emission at the position of the bridge, they should correspond to ultra-low luminosity \mgii\ emitters at $z\approx1$. 
\cite{Bacon21} has recently proposed the existence of a population of very faint $\lya$ emitters which could contribute to the extended $\lya$ emission tracing overdense structures.
We think that such a hypothesis is unlikely to hold for the \mgii\ group studied here since it would require an \mgii\ EW ($\gtrsim$20$\AA$) above the highest \mgii\ EW (i.e. 19\AA) measured in the MUSE deep fields \citep{F18}.  

{\it Shocks:} In the compact and complex configuration of our group, a large fraction of \mgii\ photons are likely to be produced due to the shocks resulting from gravitational interactions among the group members or outflows from galaxies \citep{Heckman90,Monreal06}.

In order to distinguish between these sources of ionisation, we would need additional lines like $\oiii$~$\lambda5007$ or $\rm H \beta$~$\lambda4861$ in order to use line diagnostics for a comparison with photo-ionisation and shock models (e.g. \citealt{Alarie&Morisset19}). Moreover the presence of skylines in both the \mgii\ and \oii\ doublet hampers any further investigations. 

\paragraph{Summary --}
All in all, our experiments suggests that both the UV stellar emission (at $\lambda$ < 825 $\AA$ and $\lambda$ $\approx$ 2800 $\AA$) and the UVB contribute to make the \mgii\ intragroup medium shine. The photo-ionization of Mg$^+$ by the UVB appears like the dominant scenario only if the ionizing escape fractions from the group members are close to zero, which is likely for galaxy A but questionable for galaxy C. The spatial coincidence of the \mgii\ and the non-resonant and collisionnaly excited \oii\ nebulae suggests that the in situ emission by collisions and photo-ionisation processes should be favored. The reality is likely that a mixture of different mechanisms arise at all scales. Our analysis suggests that such nebulae can be commonly found around groups of star-forming galaxies, providing that data are deep enough. This will be statistically investigated in a upcoming paper (Leclercq et al. in prep).

\subsection{Comparison with the literature}
\label{sec:63}

We now compare the properties of this new intragroup \mgii\ nebula with (i) the known ionized gas structures found around galaxy groups and (ii) recently reported \mgii\ extended emission around individual galaxies. 

\subsubsection{Ionized nebula in galaxy groups}
\label{sec:631}

Today, only a few ionized nebulae around galaxy groups (DM halo mass $\lesssim$10$^{13}$ M$_{\odot}$) are known and most of them have been detected thanks to the arrival of IFU instruments. The first detection of a large ionized structure in a galaxy group was reported by \cite{Ep18}. Using MUSE, they found a 150 kpc large \oii\ nebula at z$\simeq$0.7 embracing a dozen of galaxies with maximum stellar masses of $\approx$10$^{10.9}$ M$_{\odot}$ (the DM halo mass of the group is 6.5$\times$10$^{13}$ M$_{\odot}$, see \citealt{AbrilMel21}). By investigating the kinematics and ionisation properties, the authors concluded that gas was stripped out of the galaxies by tidal forces after galaxy interactions and AGN outflow.
A year later, \cite{Ch19} mapped a 100 kpc large H$\alpha$ structure spanning over a 14-galaxy group \citep{K10} at z$\simeq$0.3. Their analysis of the emission morphology and kinematics indicates that most of those nebulae are powered by shocks and turbulent gas motions associated with gas stripping after gravitational interactions of the group members.
Two other recent studies reported the detection of ionized nebulae emitting in \oiii, \oii\ and H$_{\beta}$ around six and three galaxy groups hosting a quasar (\citealt{Johnson18, Helton21}, respectively). 
All those studies favor a scenario where the enriched gas in the IGrM originates from tidally stripped gas after galaxy interactions.

The galaxy group studied here (i.e. embedded in the \mgii\ nebula) is at higher redshift and consists of fewer galaxies which are also less massive than the group members embedded in the other ionized nebulae. 
It also has a more compact configuration (< 50 kpc) in projection and in velocity space (< 120 km s$^{-1}$). Although the galaxy group embedded in the \mgii\ nebula shows different properties compared to the previously published systems, we all favour a scenario where the enriched gas in the IGrM originates from tidally stripped gas after galaxy interactions.

Finally, we can also mention the discovery by \cite{Rupke19} of an \oii\ nebula around a massive galaxy (10$^{11.1}$ M$_{\odot}$) at $z\approx0.46$. This nebula has an hourglass shape which is similar to our nebula. According to the authors, this spatial distribution indicates an ionized bipolar outflow. However our discovery shows that, if we have deep enough data to detect the continuum of companion galaxies, this kind of morphology is also compatible with a gas stripping scenario. Further support for this possibility comes from the two tidal tails visible in the HST image of the published \oii\ nebula.

\subsubsection{\mgii\ halos around individual galaxies}
\label{sec:632}

Thanks to the high sensitivity of MUSE and to a long exposure time, we discovered the first \mgii\ nebula around a galaxy group at any redshift. Extended \mgii\ emission around three individual galaxies at $z\simeq0.7$ has very recently been reported in \cite{Burchett21}, \cite{Zabl21} and Wisotzki et al. (in prep.) using the Keck/KCWI and MUSE instruments. These galaxies are more massive (10$^{9.75}$ < M$_*$ [M$_\odot$] < 10$^{10.05}$) than galaxy A (10$^{9.35}$ M$_\odot$) and are above the main sequence of star forming galaxies (see \citealt{Zabl21} for a compilation) which is not the case for galaxy A (Sect.~\ref{sec:32}), although the uncertainties on the SFR value are large (Table~\ref{tab:gal_prop}). Moreover, the three studies of \mgii\ halos are consistent with the presence of outflows, which is also very likely in galaxy A. 

At $z\simeq0.7$, the \oii\ emission is covered by MUSE. Both \cite{Zabl21} and Wisotzki et al. (in prep.) found extended \oii\ emission but with a steeper spatial profile compared to \mgii. The \mgii\ emission measured in these recent studies extends out from the center of the galaxies to a radius of 20 to 40 kpc. In particular, the \mgii\ halo analysed by \cite{Zabl21} spans a total area of 1000 kpc$^{2}$ above the 2$\sigma$ significance level which is comparable to our nebula. 

We can also imagine a scenario where our galaxy group would be a progenitor of the observed galaxies with \mgii\ halos at z$\simeq$0.7. In this scenario we would be seeing the pre-merging stage leading to a more massive galaxy. Actually, both the galaxies studied in Wisotzki et al. (in prep.) and \cite{Zabl21} are likely in a late stage merger as they both show significant asymmetric substructures in their central regions (see Sect.~A2 in \citealt{Zabl21}).
Using a cosmological simulation, \cite{Ventou19} investigated the probability that interacting galaxies will merge in the future. According to this study, the three galaxies A, B and C, and the two galaxies D and E, separated in projection by $\Delta$r < 25 kpc and $\Delta$v < 100 $\kms$, have 70\% chance to merge (see their Fig.~2b). The systems [A, B, C] and [D, E], separated by $\Delta$r < 50 kpc and $\Delta$v < 300 $\kms$, have at least 30\% change of merging by $z$ = 0 (their Sect.~3.2.3). 

Interestingly, both objects from \cite{Zabl21} and Wisotzki et al. (in prep.) have a less massive companion located at <100 km s$^{-1}$ in velocity space and at $\lesssim$ 50 kpc in projection. The authors do not exclude the presence of tidal stripping effects in those systems, reinforcing the idea that interactions might play a crucial role in the redistribution of metals within the circum-galactic and intragroup media. 
Finally, it appears that the signatures of outflows and tidal interactions are very difficult to disentangle. Moreover, according to simulations, we actually expect the two to co-exist, which seems to be the case in our system.

\section{Summary \& conclusions}
\label{sec:7}

Thanks to the extraordinarily deep MXDF data, we report the first detection of an intragroup medium shining in \mgii. The nebula surrounds and connects five neighboring galaxies at $z\simeq1.31$ separated by less than 50 kpc in projection and $\approx$120 km s$^{-1}$ in velocity space (Figs.~\ref{fig:group_presentation} and \ref{fig:integ_sp}). With a DM halo mass of $\approx$10$^{11.7}$ M$_{\odot}$ and highest galaxy mass of $\approx$10$^{9.3}$ M$_{\odot}$ (Table~\ref{tab:gal_prop}), this is a low mass, compact and high redshift system compared to previous groups found to be surrounded by ionised nebulae.
The detection of \mgii\ absorption features in the spectrum of a background galaxy located at an impact parameter of 19 kpc from the group indicates that the intragroup medium studied here is even larger than the nebula seen in emission.
Our $\approx$60-hour deep MUSE data (Fig.~\ref{fig:nb_mgii}b) allowed us to spatially and spectrally map the extended gaseous nebula of the system. Our observations provide a new and panoramic view on the IGrM, historically only probed one-dimensionally by studies employing absorption line techniques. This study allowed us to deepen our understanding of the existence and origin of enriched gas outside of star-forming galaxies residing in groups. Our results can be summarized as follows:
\begin{enumerate}
    \item 
    We detected a 1000 kpc$^2$ \mgii\ emitting nebula with total \mgii\ flux of 7.48$\pm$0.50 $\times$ 10$^{-18}$ \flcgs\ within a 2$\sigma$ level isophote, corresponding to a total luminosity of 8.0$\pm$0.5 $\times$ 10$^{40}$ erg s$^{-1}$. 
    Our optimized NB image construction procedure (Sect.~\ref{sec:41}) allowed us to capture most of the \mgii\ flux while limiting the noise (Appendix~\ref{ap:2}). Once slightly smoothed to increase the S/N while preserving a good spatial resolution (Appendix~\ref{ap:1}), our panoramic \mgii\ map (Fig.~\ref{fig:nb_mgii}a) reveals that the \mgii\ nebula has an elongated shape surrounding and connecting the five group members with a maximal projected extent of $\approx$70 kpc. 
    A low SB ($\approx$2$\times$10$^{-19}$ $\sbl$) bridge connects the galaxies [A,B,C] and [D,E] (Figs.~\ref{fig:nb_mgii}a and \ref{fig:sbprof_mgii}). This bridge does not appear to be a projection effect because of its tentative detection (S/N$\lesssim$2) in a single 1.25 $\AA$ (58 km s$^{-1}$) cube slice (Fig.~\ref{fig:NBlayer_mgii}).
    While the highest intensity peak coincides with galaxy C (Appendix~\ref{ap:3}), others are detected with S/N>3 within the nebula. We see clear absorption at the position of galaxy A and the background galaxy BKG (Fig.~\ref{fig:nb_mgii}c). BKG is located at an impact parameter of 19 kpc from galaxy D, outside of the 2$\sigma$ detection edge of the nebula. The detection of absorption lines in its spectrum at the redshift of the group indicates that there is gas enriched in magnesium even beyond the detected nebula. This underlines the complementarity between emission and absorption studies. 
    \item
    Taking advantage of our IFU data, we studied the kinematics of the \mgii\ gas surrounding the group using the 2D weighted Voronoi binning to increase the S/N in the lower SB regions (Sect.~\ref{sec:44}). Our resolved \mgii\ velocity map (Fig.~\ref{fig:velmap_mgii}) suggests an overall rotation of the structure along the major axis of galaxy A consistent with an extension of the ISM rotation. 
    A comparison with the \oii\ kinematics shows that the kinematics of the \mgii\ nebula is overall redshifted with respect to the \oii\ kinematics. The highest velocity variations are observed along the minor axis of galaxy A where the bulk of the gas reaches velocities of $+$120 and $-$80 km s$^{-1}$ with respect to the systemic redshift, indicative of outflowing gas.
    \item 
    At the redshift of the group, MUSE covers the \feii\ and \oii\ lines (Fig.~\ref{fig:integ_sp}). We measured extended \oii\ emission as well as a S/N$\simeq$2 \oii\ bridge at the same location as the \mgii\ bridge. The \feii\ emission appears more extended than the continuum around galaxy A and extends preferentially towards the neighboring galaxy D, similarly to the \mgii\ and \oii\ emission (Fig.~\ref{fig:oiifeii_nbsb}).
    \item 
    This first panoramic and 3D view of an IGrM detected in \mgii\ revealed a diffuse low SB metal-enriched gaseous component surrounding and connecting the galaxy group members (Sect.~\ref{sec:61}). The evidence for outflows, at least from the most massive galaxy A, as well as the proximity of the five group members, suggest that the IGrM has been enriched through both stellar outflows and tidal interactions between the group members (Sect.~\ref{sec:621}). 
    \item
    We listed the different mechanisms that can power the \mgii\ nebula in Sect.~\ref{sec:622}: stellar UV continuum at \mgii, nebular emission, collisional excitation, and discussed the possible distributions of sources that can explain the intragroup \mgii\ luminosity and light distribution.
    We found that the UV stellar continua, both at $\lambda$ < 825 $\AA$ and $\lambda$ $\approx$ 2800 $\AA$, are possible sources powering the observed \mgii\ nebula (Appendix~\ref{ap:51}). We compare the stellar emission to the UVB contribution (Appendix~\ref{ap:52}) and found that the UVB is the dominant process only if the total ionizing escape fraction is close to zero which is likely the case for galaxy A, but questionable for galaxy C.
    The spatial coincidence of the \mgii\ and \oii\ nebulae favours the in situ emission by collisional and photo-ionisation processes. We conclude that we need additional lines to distinguish between the different sources of ionization/excitation.
    \item
    When comparing our results with the literature (Sect.~\ref{sec:63}) and specifically to previously detected ionized nebulae in galaxy groups (DM halo mass $\lesssim$10$^{13}$ M$_{\odot}$), we found that our nebula is less extended and surrounds fewer and lower mass galaxies. Moreover, a comparison between our \mgii\ nebula and the three \mgii\ halos recently mapped by IFUs and detected around individual galaxies at lower redshift ($z\simeq0.7$), reveals similar spatial extents. For those three cases, the authors reported the presence of outflows. Interestingly, two of the three galaxies from the literature have a close companion galaxy implying that tidal stripping cannot be excluded. Finally, we suggest the possibility that our $z=1.3$ system constitutes a pre-merging stage of the observed $z\simeq0.7$ galaxies surrounded by an \mgii\ halo. This is reinforced by the fact that all three $z\simeq0.7$ galaxies show hints for a late stage merger.
\end{enumerate}

This discovery paper shed light on the existence and origin of the IGrM in one low mass system of five galaxies. More observations are needed to generalize the presence of a low SB diffuse enriched gaseous component within low mass galaxy groups. While facilitated by the advent of IFU instruments like MUSE, such observations are however very expensive in telescope time as they require several tens of hours in exposure time. More cases of extended $\mgii$ emission have been detected in the MXDF data and are the topic of an upcoming paper (Leclercq et al. in prep.). 

Using the \mgii\ emission to map the circumgalactic and intragroup media is particularly interesting because it traces the same cool and neutral gas phase as the $\lya$ emission. By comparing $\lya$ and \mgii\ halo properties we will be able to probe the spatial distribution of the enriched versus pristine gas around galaxies and better characterize the gas exchanges between the galaxies and their environments, which is crucial to understand galaxy evolution.


\begin{acknowledgements}
F.L. and A.V. acknowledge support from SNF Professorship PP00P2\_176808.
A.V. and T.G. are supported by the ERC starting grant ERC-757258-TRIPLE. This work is partly funded by Vici grant 639.043.409 from the Dutch Research Council (NWO). SC gratefully acknowledges support from Swiss National Science Foundation grant PP00P2\_190092 and from the European Research Council (ERC) under the European Union's Horizon 2020 research and innovation programme grant agreement No 864361. NB acknowledges support from the ANR 3DGasFlows (ANR-17-CE31-0017).
\end{acknowledgements}


\bibliographystyle{aa} 
\bibliography{biblio}

\begin{appendix}

\onecolumn

\section{Nebula appearance at different depths}
\label{ap:4}

\begin{figure*}[h!]
\centering
   \resizebox{\hsize}{!}{\includegraphics{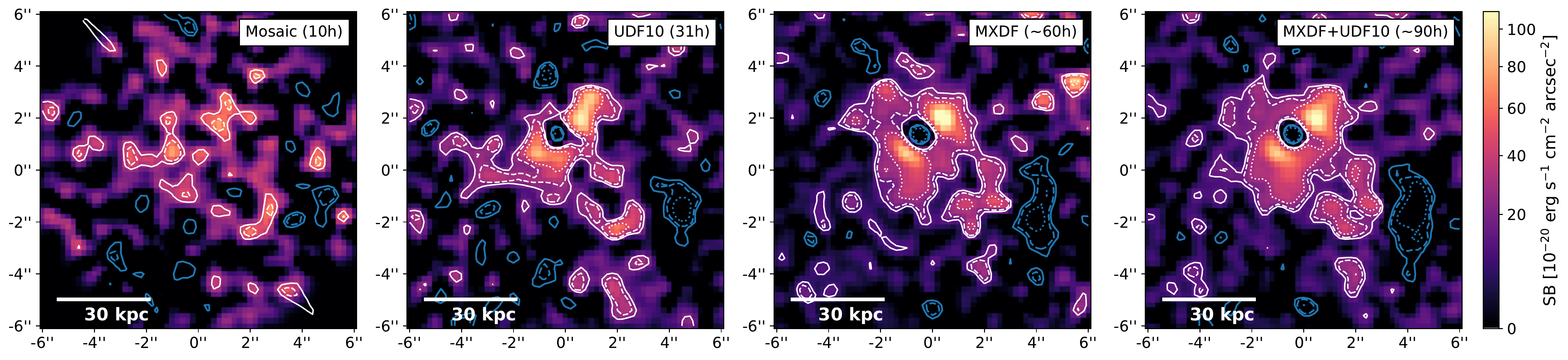}}
    \caption{\mgii\ NB image of the nebula as constructed in Sect.~\ref{sec:41} using the \mosaic\ (10h), \udft\ (31h), MXDF ($\approx$60h) and a combination of the \udft\ and MXDF ($\approx$90h) MUSE data cubes. The legend is the same as in Fig.~\ref{fig:nb_mgii}a.}
    \label{fig:neb_udf10_mxdf}
\end{figure*}

The \mgii\ nebula was first discovered in the 31-hour deep MUSE \udft\ data cube (second panel of Fig.~\ref{fig:neb_udf10_mxdf}). Part of this field has been re-observed at greater depth ($\approx$60 h) as part of the MXDF program (Sect~\ref{sec:2}). Here we compare the appearance of the detected \mgii\ IGrM in these two fields (left and middle panels) and in the combined datacube (\udft\ and MXDF) reaching a $\approx$90 h depth (right panel). 

In the combined data, the shape of the \mgii\ nebula is reassuringly preserved, the S/N increased but the PSF is larger. For comparison we also show the appearance of the nebula in the 10-hour deep \mosaic\ field, highlighting the fact that deep observations are crucial to study the CGM and IGrM of low mass galaxies. 

\section{Smoothing effects on the detected nebula shape}
\label{ap:1}

Although our MUSE data are very deep ($\approx$60 h, see Fig.~\ref{fig:nb_mgii}b), the diffuse \mgii\ emission of the group revealed by such deep observations is still very faint.
In order to increase the S/N while keeping an as good as possible spatial resolution, we smoothed our data with a Gaussian kernel whose FWHM (3 MUSE pixels or 0\farcs6) is slightly higher than the MUSE PSF FWHM (0\farcs52). 

\begin{figure*}[h!]
\centering
   \resizebox{\hsize}{!}{\includegraphics{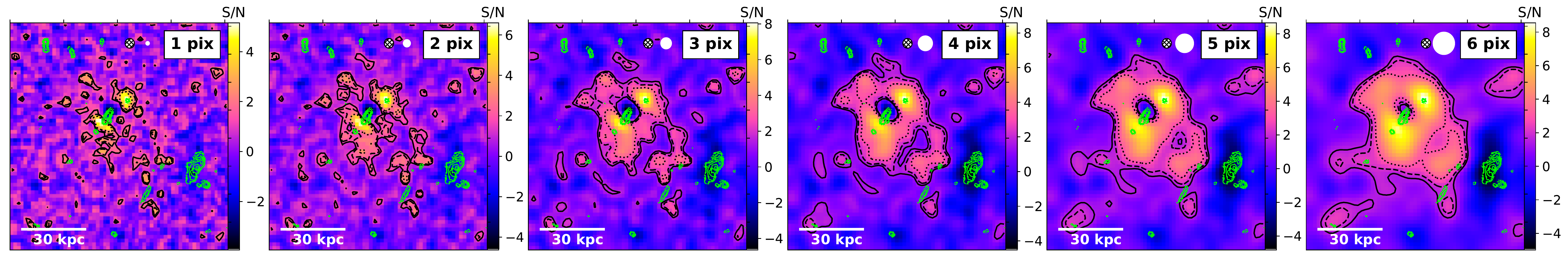}}
    \caption{\mgii\ S/N map as constructed in Sect.~\ref{sec:41} and smoothed using a 1, 2, 3, 4, 5 and 6 spaxel FWHM Gaussian from left to right, respectively. The FWHM values are indicated and shown as white circle (top right) in each panel. For comparison, the FWHM of the MUSE PSF is shown with a hatched white circle. The solid, dashed and dotted black contours correspond to a S/N of 1.5, 2 and 3, respectively. 
    The adopted smoothing kernel has a FWHM of 0\farcs6 (i.e. 3 spaxels or $\approx$5 kpc at $z=1.31$) and allows to improve the S/N while keeping a good spatial resolution.}
    \label{fig:snrmap_smo}
\end{figure*}

Figure~\ref{fig:snrmap_smo} shows the dependence of the \mgii\ nebula S/N map on the smoothing kernel size. We choose to apply a smoothing as light as possible to keep an as good as possible spatial resolution.

\section{Spectral bandwidth effect on the \mgii\ nebula detection}
\label{ap:2}

\begin{figure*}[h!]
\centering
   \resizebox{\hsize}{!}{\includegraphics{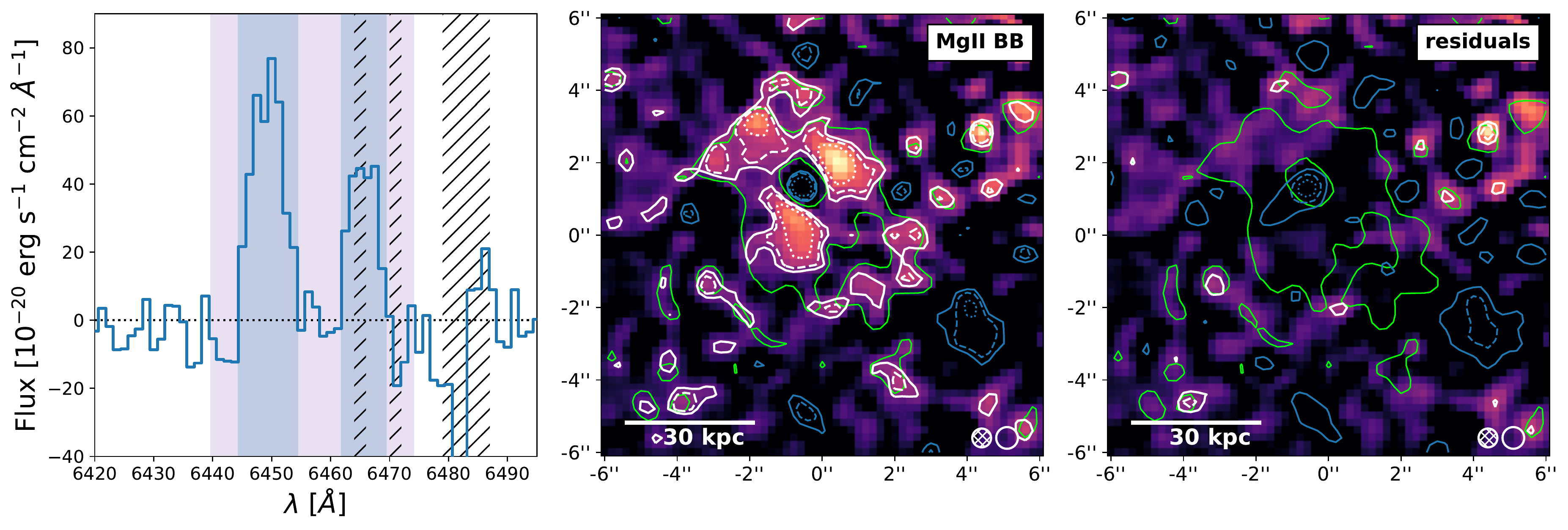}}
    \caption{\textit{Left:} \mgii\ spectrum of the nebula integrated within the 2$\sigma$ significance level contour (see Fig.~\ref{fig:nb_mgii}a). The vertical blue shaded areas indicate the spectral widths used to construct the \mgii\ NB image shown in Fig.~\ref{fig:nb_mgii}. The shaded purple area shows the wavelength band used to build the broader NB image shown in the middle panel. The hatched areas indicate the position of sky lines.
    \textit{Middle:} A broader $\mgii$ NB image extended by $\pm$500 km s$^{-1}$ on the $\lambda$2796 and $\lambda$2803 line outskirts (purple shaded area in the left panel) encompassing the whole \mgii\ spectral range (2792$-$2806$\AA$ in rest frame). 
    \textit{Right:} Difference between the \mgii\ NB and broader band (blue and purple shaded areas in the left panel) images. The contours are the same as in Fig.~\ref{fig:nb_mgii} except that the green contour here shows the 1.5$\sigma$ contour of the S/N-optimized \mgii\ NB image.}
    \label{fig:nb-vs-bb}
\end{figure*}

As described in Sect.~\ref{sec:41}, the final \mgii\ NB image (Fig.~\ref{fig:nb_mgii}a) has been optimized to encompass all the \mgii\ flux in the $\lambda$2796 and $\lambda$2803 lines while limiting the noise. The selected wavelength bands are shown by the shaded blue areas in the first panel of Fig.~\ref{fig:nb-vs-bb} and indeed cover both the \mgii\ lines extracted in an area above the 2$\sigma$ significance level (first panel). In order to emphasize the importance of optimizing the NB image and to check for any missing flux, we created a broader \mgii\ NB image (middle panel) encompassing the whole \mgii\ spectral range (2792$-$2806$\AA$ in rest frame, purple shaded area in left panel). This broader-band (BB) \mgii\ image is noisier and the overall shape (green contour) of the nebula is lost in the noise. Looking at the residual image (i.e. the difference between the NB and BB \mgii\ images, right panel), we can see that, while some absorbing flux at the position of galaxy A is missed, all the emitting flux is captured by our NB construction procedure.

\section{Optically thin emission in the galaxy C line of sight}
\label{ap:3}

\begin{figure}[h!]
\centering
   \resizebox{0.5\hsize}{!}{\includegraphics{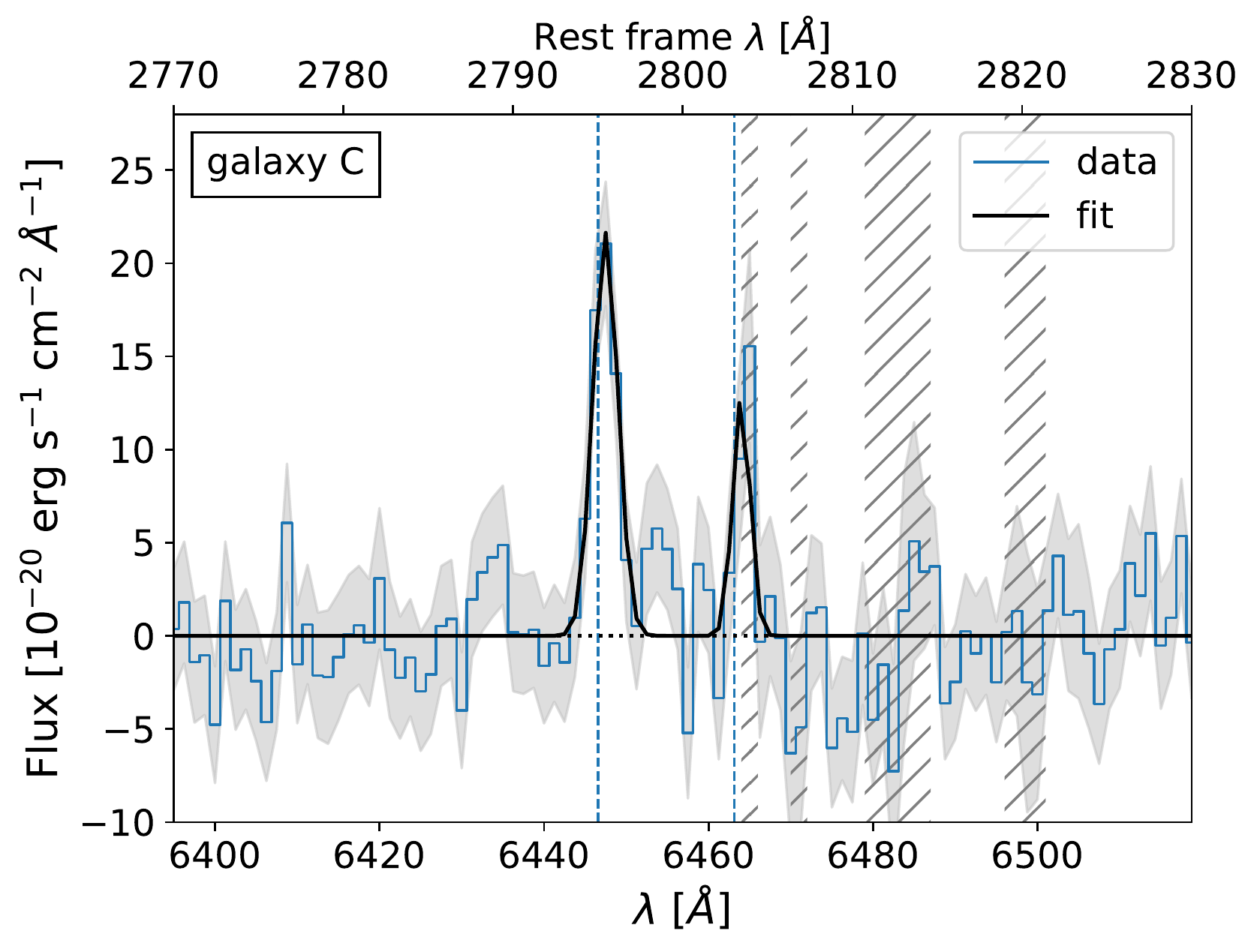}}
    \caption{\mgii\ doublet line fitting of galaxy C. The best fit line (black) is shown along with the continuum subtracted \textsf{ODHIN} spectrum (blue). The uncertainties are shown in grey and the position of sky lines with grey hatches.}
    \label{fig:C_mgii_fit}
\end{figure}

In Sect.~\ref{sec:43} we rescaled the continuum radial SB profile to the \mgii\ radial SB value matching the position of galaxy C. This choice is motivated by several reasons: 
\begin{itemize}
    \item 
    the fact that the lines are well fitted by Gaussian functions (see Fig.~\ref{fig:C_mgii_fit}) indicates that the \mgii\ light emitted from galaxy C does not undergo strong radiative transfer (RT) effects or absorption
    \item
    the best-fit \mgii\ line flux ratio F$_{2796}$/F$_{2803}$ of 2.11$\pm$0.69 for galaxy C corresponds to optically thin emission and thus is in good agreement with a lack of RT effects and absorption in this line of sight 
    \item 
    as a resonant line, we can expect the so-called infilling effect to be significant for \mgii\ in this system. In other words, the absorbed photons can be re-emitted resonantly and escape the galaxy. As such, scattered \mgii\ photons can contribute to the spectrum of the galaxy by "filling" the \mgii\ absorption troughs. \cite{Mauerhofer21} studied this effect for a simulated galaxy and found that it is more important for directions showing strong fluorescence lines. No fluorescent \feii\ lines are significantly detected in the galaxy C spectrum (Fig.~\ref{fig:integ_sp}, third line). The lack of infilling would reinforce the lack of scattering in the line of sight of galaxy C because the photons creating the infilling effect must have scattered at least once according to \cite{Mauerhofer21}.
\end{itemize}
To sum up, the lack of hints for absorption features and RT effects in the spectrum of galaxy C indicate that the \mgii\ emission is optically thin at the location of C. Those motivations justify our continuum rescaling procedure used to highlight the extended nature of the \mgii\ emission on radial SB profiles (Sect.~\ref{sec:43}).

\section{Estimating the intrinsic \mgii\ photons budget using \textsf{Cloudy} modelling}
\label{ap:5}

\subsection{Intrinsic nebular emission in \hii\ regions}
\label{ap:51}

In order to predict an \mgii\ intrinsic stellar photon budget, we produced a grid of \textsf{Cloudy} models with a closed spherical geometry \citep{Ferland2017}. The input SEDs were created by combining the \textsf{MAGPHYS} intrinsic spectra for the group galaxies (obtained by fitting the \citealt{R15} HST photometry, see Sect~\ref{sec:32}) with constant star forming model SEDs obtained from the binary population synthesis code \textsf{BPASS}v2.2.1 \citep{Eldridge2017} as shown in Fig.~\ref{fig:UVB}. The \textsf{BPASS} SED was used to extend the \textsf{MAGPHYS} spectra to ionising wavelengths. The age of the stellar population models was set to 10$^{8.75}$~years and the metallicity ($Z$) to two times solar, in accordance to the \textsf{MAGPHYS} best fit parameters. The stopping criteria of the calculations were set by the Hydrogen column density, with values between log$_{10}$($N_{\rm{Hstop}}$) = 17 and 20 cm$^{-2}$. This parameter cuts the calculation at some column density values inside the Stömgren Sphere.
Finally, the luminosity of the models was set by the amount of hydrogen ionising photons ($\log(q(H)) \simeq 51$ photons s$^{-1}$), constraining the ionisation parameter to $\log<U> \simeq -2.6$.

We adopt two different prescriptions for the element abundances in the cloud: (1) solar abundance ratios from \cite{Grevesse2010} where $\frac{Mg}{H} \sim 10^{-4.5}$ and, (2) ISM abundances based on the mean warm and cold phases of the ISM ($\frac{Mg}{H} \sim 10^{-4.9}$). It is noteworthy to mention that (1) does not include grains, while (2) has a combination of silicates and graphites defined by \cite{Mathis1977}.  The results of these models can be seen in Figure~\ref{fig:cloudy}, in which we show how the intrinsic \mgii\ to \oii\ ratio changes as a function of the stopping column density. We find that the models provide a higher intrinsic \mgii\ to \oii\ ratio than the observed one (where dust attenuation and scattering might explain the discrepancy) when considering solar abundance ratios and ISM abundances with a stopping column density > 10$^{18}$ cm$^{-2}$. This suggests that the \mgii\ emission of the nebula can be explained by stellar processes alone through resonant scattering of photons originally produced in \hii\ regions. 

\subsection{Comparison between the stellar versus UVB contribution to the \mgii\ nebulae}
\label{ap:52}

In order to assess the origin of the Mg$^+$ ionizing photons ($\lambda$ < 825 $\AA$), leading to the emission of \mgii\ photons, we compare the total SED of the group members (sum of the five SEDs, see Sect.~\ref{ap:51}) with the UVB SED from \cite{HaardtMadau12} at the redshift of the group (i.e. $z$ = 1.3). Those SEDs are in rest frame and show the flux as it would be observed at a distance varying between 10 and 35 kpc of the stars. This range in radius allowed us to evaluate and compare the contributions of the stars and UVB at the edge of the detected nebula as well as closer to the galaxies. We show two models with arbitrary values of the escape fractions 0.7\% and 14\% (log$_{10}$($N_{\rm{Hstop}}$) = 18 and 18.5, respectively), calculated as the transmitted to intrinsic ratio of the flux integrated < 912 $\AA$. 

\begin{figure}[h!]
   \begin{minipage}[c]{.46\linewidth}
      \resizebox{0.97\hsize}{!}{\includegraphics{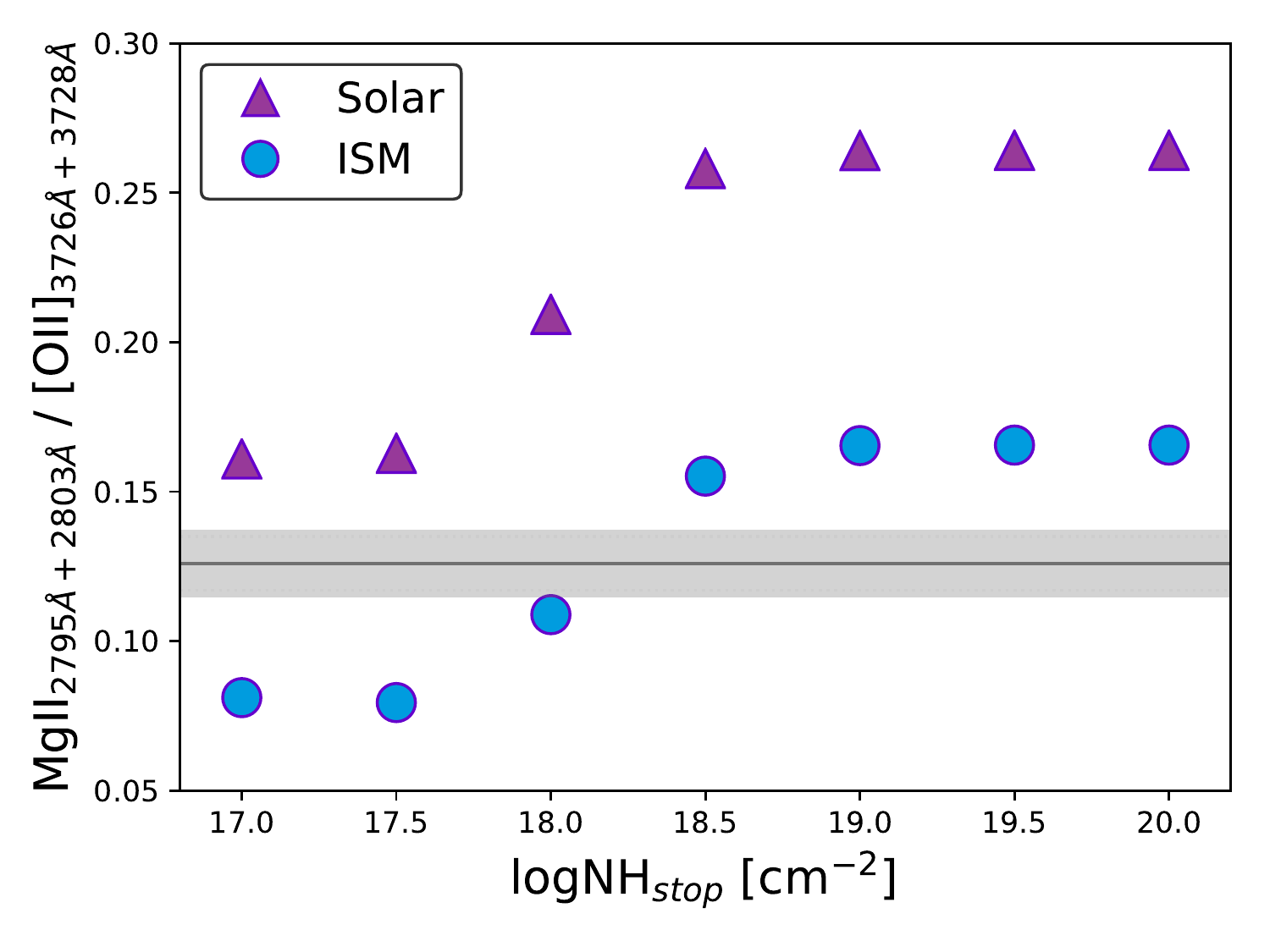}}
       \caption{Predicted intrinsic nebular \mgii\ to \oii\ ratio using \textsf{Cloudy} models constrained by the best fit parameters of the observations (see SEDs in the right panel). The symbols represent the chemical abundances used, triangles for solar and circles for ISM. The horizontal grey line and shaded area show the observed \mgii\ to \oii\ ratio from the group and the 1$\sigma$ uncertainties, respectively.}
       \label{fig:cloudy}
   \end{minipage} \hfill
   \begin{minipage}[c]{.52\linewidth}
      \resizebox{0.97\hsize}{!}{\includegraphics{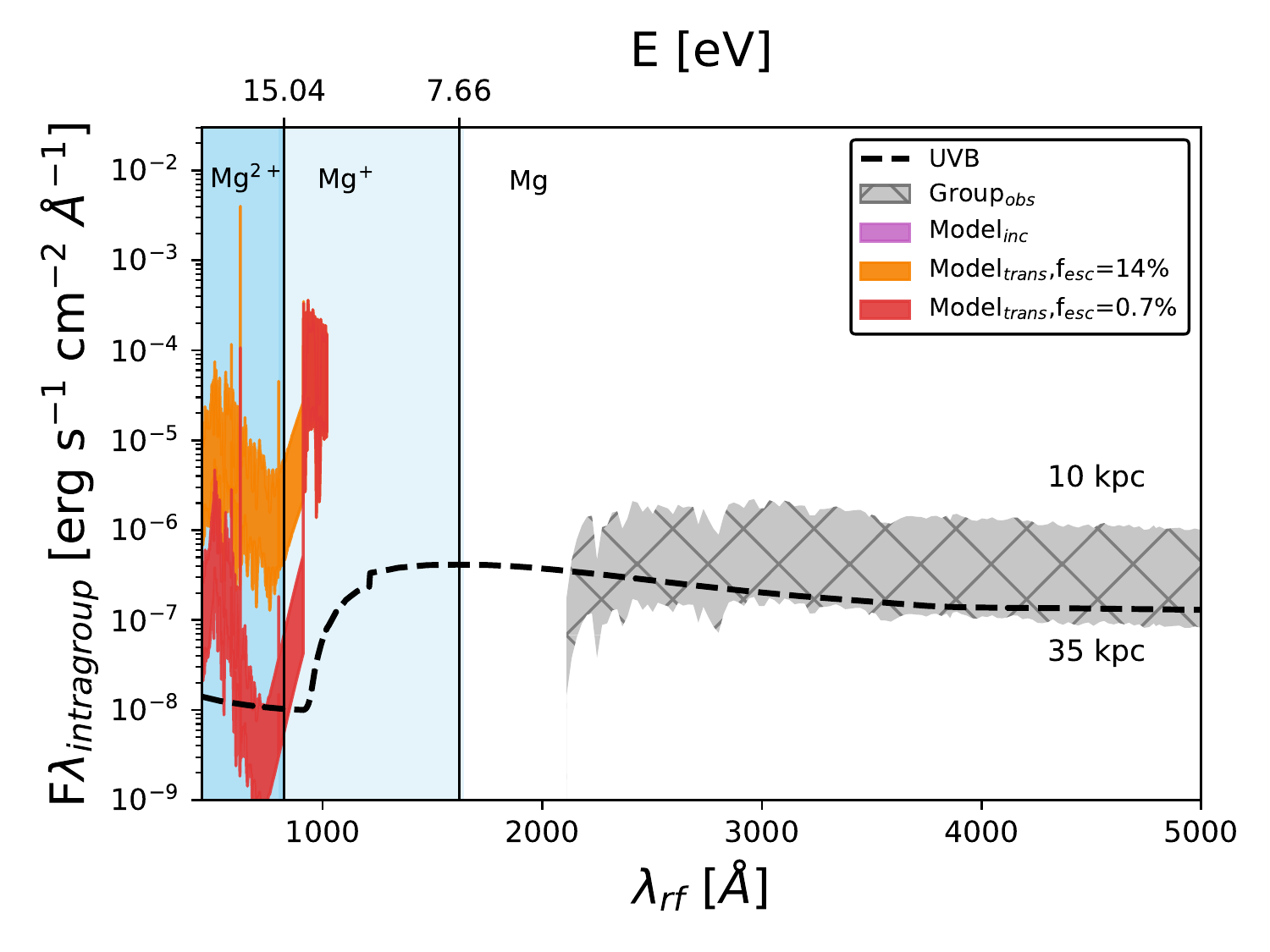}}
      \caption{Comparison of the UVB vs stellar photons budget, ionizing Mg$^{+}$ in Mg$^{2+}$. In purple we show the intrinsic \textsf{BPASS} model for $Z = 2~Z_\odot$ and age of 10$^8.75$ years for a constant SF stellar population.
      The transmitted \textsf{BPASS} model considering 0.7\% and 14\% ionizing escape fraction is shown in red and orange, respectively.
      The grey spectrum shows the attenuated spectrum of the group (sum of five SEDs, Sect.~\ref{ap:51}). The dispersion of all spectra depends on the radius at which we estimate the flux, the lower and upper boundaries of the areas are for 35 and 10 kpc, respectively. The black dashed line shows the UVB from \cite{HaardtMadau12} at $z$ = 1.3.}
    \label{fig:UVB}
   \end{minipage}
\end{figure}

In Figure~\ref{fig:UVB} we show that the contribution of the UVB to the Mg$^+$ ionizing flux ($\lambda$<825 $\AA$ or E>15.04 eV) is dominant in the nebula (10 < $r$ [kpc] < 35) only when considering an ionizing escape fraction close to zero. When getting closer to the stars ($r$~<~10~kpc), the contribution of the UV stellar emission becomes dominant.
At higher escape fractions (e.g. 14\% as shown on the figure), the contribution of the UVB quickly becomes negligible, even at the outskirts of the nebula.

\end{appendix}

\end{document}